\PassOptionsToPackage{unicode}{hyperref}
\PassOptionsToPackage{hyphens}{url}
\PassOptionsToPackage{dvipsnames,svgnames,x11names}{xcolor}
\documentclass[12pt]{article}

\usepackage{amsmath,amssymb}
\usepackage{iftex}
\ifPDFTeX
  \usepackage[T1]{fontenc}
  \usepackage[utf8]{inputenc}
  \usepackage{textcomp} 
\else 
  \usepackage{unicode-math}
  \defaultfontfeatures{Scale=MatchLowercase}
  \defaultfontfeatures[\rmfamily]{Ligatures=TeX,Scale=1}
\fi
\usepackage{placeins}
\usepackage{lmodern}
\usepackage{amsmath,amssymb} 
\usepackage{mathrsfs}
\usepackage{setspace}
\usepackage{amsthm} 
\newcommand{\cS}{\mathcal{S}} 
\ifPDFTeX\else  
\fi

\numberwithin{equation}{section}

\theoremstyle{plain}          
\newtheorem{assumption}{Assumption}[section]
\newtheorem{lemma}{Lemma}[section]
\newtheorem{theorem}{Theorem}[section]

\newtheorem{remark}{Remark}[section]   

\IfFileExists{upquote.sty}{\usepackage{upquote}}{}
\IfFileExists{microtype.sty}{
  \usepackage[]{microtype}
  \UseMicrotypeSet[protrusion]{basicmath} 
}{}
\makeatletter
\@ifundefined{KOMAClassName}{
  \IfFileExists{parskip.sty}{%
    \usepackage{parskip}
  }{
    \setlength{\parindent}{0pt}
    \setlength{\parskip}{6pt plus 2pt minus 1pt}}
}{
  \KOMAoptions{parskip=half}}
\makeatother
\usepackage{xcolor}
\makeatletter
\ifx\paragraph\undefined\else
  \let\oldparagraph\paragraph
  \renewcommand{\paragraph}{
    \@ifstar
      \xxxParagraphStar
      \xxxParagraphNoStar
  }
  \newcommand{\xxxParagraphStar}[1]{\oldparagraph*{#1}\mbox{}}
  \newcommand{\xxxParagraphNoStar}[1]{\oldparagraph{#1}\mbox{}}
\fi
\ifx\subparagraph\undefined\else
  \let\oldsubparagraph\subparagraph
  \renewcommand{\subparagraph}{
    \@ifstar
      \xxxSubParagraphStar
      \xxxSubParagraphNoStar
  }
  \newcommand{\xxxSubParagraphStar}[1]{\oldsubparagraph*{#1}\mbox{}}
  \newcommand{\xxxSubParagraphNoStar}[1]{\oldsubparagraph{#1}\mbox{}}
\fi
\makeatother

\usepackage{longtable,booktabs,array}
\usepackage{calc} 
\usepackage{etoolbox}
\usepackage[ruled,vlined]{algorithm2e} 
\numberwithin{equation}{section}
\numberwithin{figure}{section}
\numberwithin{table}{section}

\SetKwInput{KwInput}{Input}           
\SetKwInput{KwOutput}{Output}         

\usepackage{enumitem}

\makeatletter
\patchcmd\longtable{\par}{\if@noskipsec\mbox{}\fi\par}{}{}
\makeatother
\IfFileExists{footnotehyper.sty}{\usepackage{footnotehyper}}{\usepackage{footnote}}
\makesavenoteenv{longtable}
\usepackage{graphicx}
\makeatletter
\def\maxwidth{\ifdim\Gin@nat@width>\linewidth\linewidth\else\Gin@nat@width\fi}
\def\maxheight{\ifdim\Gin@nat@height>\textheight\textheight\else\Gin@nat@height\fi}
\makeatother
\setkeys{Gin}{width=\maxwidth,height=\maxheight,keepaspectratio}
\makeatletter
\def\fps@figure{htbp}
\makeatother

\makeatletter
\@ifpackageloaded{caption}{}{\usepackage{caption}}
\AtBeginDocument{%
\ifdefined\contentsname
  \renewcommand*\contentsname{Table of contents}
\else
  \newcommand\contentsname{Table of contents}
\fi
\ifdefined\listfigurename
  \renewcommand*\listfigurename{List of Figures}
\else
  \newcommand\listfigurename{List of Figures}
\fi
\ifdefined\listtablename
  \renewcommand*\listtablename{List of Tables}
\else
  \newcommand\listtablename{List of Tables}
\fi
\ifdefined\figurename
  \renewcommand*\figurename{Figure}
\else
  \newcommand\figurename{Figure}
\fi
\ifdefined\tablename
  \renewcommand*\tablename{Table}
\else
  \newcommand\tablename{Table}
\fi
}
\@ifpackageloaded{float}{}{\usepackage{float}}
\floatstyle{ruled}
\@ifundefined{c@chapter}{\newfloat{codelisting}{h}{lop}}{\newfloat{codelisting}{h}{lop}[chapter]}
\floatname{codelisting}{Listing}

\makeatother
\makeatletter
\@ifpackageloaded{caption}{}{\usepackage{caption}}
\@ifpackageloaded{subcaption}{}{\usepackage{subcaption}}
\makeatother

\ifLuaTeX
  \usepackage{selnolig}  
\fi
\usepackage[]{natbib}
\usepackage{bookmark}

\IfFileExists{xurl.sty}{\usepackage{xurl}}{} 
\hypersetup{
  pdftitle={Title},
  pdfauthor={Author 1; Author 2},
  pdfkeywords={3 to 6 keywords, that do not appear in the title},
  colorlinks=true,
  linkcolor={blue},
  filecolor={Maroon},
  citecolor={Blue},
  urlcolor={Blue},
  pdfcreator={LaTeX via pandoc}}

\newcommand{\anon}{1}

\begin{document}

\def\spacingset#1{\renewcommand{\baselinestretch}%
{#1}\small\normalsize} \spacingset{1}


\def\anon{1}  
\if1\anon
{
  \title{\bf Bootstrapped Control Limits
        for Score-Based Concept Drift Control Charts}
  \author{Jiezhong Wu\\
    Department of Industrial Engineering and Management Sciences,\\ Northwestern University\\
    and \\
    Daniel W. Apley\\
    Department of Industrial Engineering and Management Sciences, \\Northwestern University}
  \maketitle
} \fi

\if0\anon
{
  \bigskip
  \bigskip
  \bigskip
  \begin{center}
    {\LARGE\bf Bootstrapped Control Limits
        for Score-Based Concept Drift Control Charts}
\end{center}
  \medskip
} \fi

\bigskip
\begin{abstract}
Monitoring for changes in a predictive relationship represented by a fitted supervised 
learning model (i.e., concept drift detection) is a widespread problem in modern 
data-driven applications. A general and powerful Fisher score–based concept drift 
approach was recently proposed by \citet{Zhang2023}, in which detecting concept drift 
reduces to detecting changes in the mean of the model's score vector using a multivariate 
exponentially weighted moving average (MEWMA). To implement the approach, the initial 
data must be split into two subsets. The first subset serves as the training sample to 
which the model is fit, and the second subset serves as an out-of-sample test set from 
which the MEWMA control limit (CL) is determined.
{In this paper, we retain the same score-based MEWMA monitoring statistic 
as \citet{Zhang2023} and develop a novel nested bootstrap procedure computing a much more 
accurate CL that also allows the entire initial data to be used for model fitting, thereby 
yielding a more accurate baseline model while eliminating the need for a large holdout set. 
The outer bootstrap loop is fully parallelizable, making the method computationally 
practical, with CL setup times comparable to or faster than those of \citet{Zhang2023}.} 
We show that a standard nested bootstrap substantially underestimates the variability of 
the monitoring statistic and develop a 0.632-like correction that appropriately accounts 
for this. We demonstrate the advantages with numerical examples.
\end{abstract}

\noindent%

\spacingset{1.8} 

\section{Introduction}
The increasing reliance on data-driven decision making has led to the widespread adoption of supervised learning models across various domains. These models aim to capture the predictive relationship $\mathbb{P}(Y \vert \mathbf{X})$ between a response variable $Y$ and covariates $\mathbf{X}$. However, a fundamental challenge arises when the predictive relationship in new data deviates from that used to train the model, potentially rendering the model's predictions unreliable or obsolete \citep{Webb2016, Zhang2023, Malinovskaya2023}, and/or reflecting a change in process behavior that should be detected. This challenge is particularly acute in domains like finance \citep{Sun2017} and healthcare analytics \citep{Razak2023}, as most AI systems and algorithms require training data that may contain inherent biases or may not remain representative of the broader population over time \citep{Barocas2019}.

Statistical process monitoring (SPM) and control charts have long served as foundational tools for detecting changes in process characteristics over time \citep{Montgomery2020}. The field has evolved considerably, with recent advances incorporating machine learning methods like kernel methods \citep{Apsemidis2020} and neural networks \citep{Psarakis2011} to enhance monitoring capabilities. As datasets have grown larger and organizations increasingly rely on predictive models, the predictive relationship between variables has emerged as a critical characteristic that requires monitoring for quality control purposes, alongside traditional process variables monitoring.

This evolution in the goals of monitoring reflects a fundamental shift in how data are used in practice and/or represents a change in process behavior that should be detected. While traditional SPM methods are designed to detect shifts in the distribution of process variables, modern applications often require understanding changes in the predictive relationships themselves. This phenomenon, known as concept drift in the machine learning literature, occurs when the relationship between input features $\mathbf{X}$ and the target variable $Y$ evolves over time \citep{Webb2016}, potentially degrading model performance \citep{MorenoTorres2012, Webb2016, Zliobaite2016} and/or representing a change in process behavior that should be detected. These changes can manifest gradually or abruptly, and may not affect predictive accuracy immediately \citep{Gama2014}, making them difficult to detect using conventional methods.

Most existing concept drift detection methods fall into two categories, with each having significant limitations. The first category consists of error-based approaches that rely primarily on monitoring classification error rates or prediction accuracy metrics \citep{Baena2006, Ross2012}. While straightforward to implement, these methods often fail to detect concept drift when changes in $\mathbb{P}(Y \vert \mathbf{X})$ do not manifest as increased error rates, such as when decision boundaries shift in ways that maintain similar overall accuracy despite fundamental changes in the underlying relationship. The second category comprises adaptive learning algorithms, which continually retrain or update models to adapt to incoming data streams \citep{Wang2003, Gama2014, Krawczyk2017}. While these approaches can improve responsiveness, the adaptive model retraining also requires significant computation and may overfit transient changes. 

Related to, but distinctly different from, concept drift detection, profile monitoring methods in statistical process control have been extensively developed to monitor functional relationships between response and predictor variables \citep{woodall2004using}. 
These methods typically involve monitoring regression parameters or fitted curves over time to detect changes in the relationship between a response variable and one or more explanatory variables \citep{Chang2010, Abbasi2022}. 
{A profile in this context refers to the fitted functional relationship (e.g., a regression curve or model) that characterizes how $Y$ depends on $\mathbf{X}$ within a given batch of data.
While this may appear similar to concept drift detection, profile monitoring fundamentally differs in its objectives, its approaches, and the structure of the data to which it applies.}
In profile monitoring, data are grouped into batches of $(\mathbf{X}, Y)$ observations, where some feature of $\mathbb{P}(Y|\mathbf{X})$ (typically $\mathbb{E}[Y|\mathbf{X}]$) for each batch represents a profile associated with the batch, and the objective is to monitor for changes in the nature of the profiles from batch to batch. 
This requires fitting predictive models separately to each batch of data. 
In contrast, in concept drift detection, data consist of individual, sequentially arriving $(\mathbf{X}, Y)$ observations, and the objective is to detect whether the predictive relationship for new observations changes relative to what it was when a baseline model was fit to prior training data. 
Changes in the predictive relationship are detected by comparing a new stream of individual $(\mathbf{X}, Y)$ observations to the baseline model, which does not involve fitting separate models to the new data. 

Recently, \citet{Zhang2023} introduced a new Fisher score based concept drift detection that uses well-established statistical theory to show that detecting changes in $\mathbb{P}(Y|\mathbf{X})$ is equivalent to detecting changes in the mean of the score vector (the gradient of the log-likelihood) of the supervised learning model, a more familiar problem in SPM for which a conventional multivariate exponentially weighted moving average (MEWMA) control charts can be used. Their approach demonstrates superior detection power compared to traditional error-based concept drift methods, while also providing valuable diagnostic information to help understand the nature of detected changes. It also avoids having to continuously retrain the model, as in adaptive learning algorithms. Computing the MEWMA control limit (CL) in \citet{Zhang2023} requires that a second large sample of ($\mathbf{X}$, $Y$) observations be collected, in addition to the training sample to which the baseline model is fit. This can be achieved by splitting the training sample into two sets (with the baseline model fit to the first set, and the second set used to compute the CL), but this reduces the size of the data to which the baseline model is fit and generally requires very large sample sizes to accurately control false-alarm rate. 
{This approach also shares conceptual similarities with risk-adjusted monitoring in healthcare \citep{Steiner2000}, which monitors residuals from a risk model (typically a GLM) and essentially falls into the same category as the error-based concept drift methods discussed above.}

{Our main contribution is that we present a more data-efficient and reliable approach that uses bootstrapping to compute the CL from the same training sample to which the baseline model is fit, thereby allowing the entire sample to be used to train a more accurate baseline model, as well as providing much more accurate control of false-alarm rate.} For this we develop a nested bootstrap procedure with the inner loop accounting for future data variability and outer loop accounting for training sample variability. This is challenging because a naïve nested bootstrap procedure substantially underestimates variability of the MEWMA monitoring statistic for reasons that we discuss later. 
{To account for this, we develop a 0.632-like (analogous to, but different from the original 0.632 correction of \citet{Efron1983}) that controls false-alarm rate much more accurately than the two-sample CL calculation of \citet{Zhang2023}, especially when the training sample size and/or the desired false-alarm rate is small.} 
Our implementation takes advantage of the inherent parallelism in the nested bootstrap structure, making the method computationally efficient and practical for modern applications involving complex models like deep neural networks. {As shown in Appendix~\ref{app3} (Table~\ref{tab:compute_compare}), with modest parallelization 
of the outer bootstrap loop the CL setup time is comparable to or faster than the time 
required by \citet{Zhang2023}.}

{We note that the detection statistic utilized in our approach is identical to the score-based MEWMA statistic proposed by \citet{Zhang2023}. Extensive empirical comparisons in \citet{Zhang2023} have already demonstrated that this statistic yields superior detection power compared to error-based methods (e.g., DDM \citep{Gama2004}, EDDM \citep{Baena2006}) and adaptive windowing, particularly for drifts that do not immediately degrade prediction accuracy. Therefore, our numerical analyses specifically on the accuracy of the proposed bootstrap Control Limits (Type I error control) and data efficiency.}

The remainder of the paper is organized as follows.
Section~\ref{Score Vectors and Concept Drift Monitoring} reviews the score-based concept drift detection approach that is the basis of our approach. Section~\ref{Bias Detection Algorithms and Derivations} develops a nested bootstrap approach and derives the associated variance-inflation correction that leads to appropriate CL. Section~\ref{Numerical Illustrations} demonstrates
our method’s effectiveness through a variety of examples ranging from linear mixture models to complex nonlinear dynamical systems. Section~\ref{conclude} concludes with a summary of the main findings in this paper.

\section{Background on Score-Based Concept Drift Detection}\label{Score Vectors and Concept Drift Monitoring}

This work assumes a parametric supervised learning $g(\boldsymbol{\theta};\mathbf{X})$ is used to represent the conditional distribution $\mathbb{P}(Y \vert \mathbf{X}; \boldsymbol{\theta})$ of the target variable $Y$ given the input features $\mathbf{X}\in\mathbb{R}^p$, where $\boldsymbol{\theta}$ denotes the parameters of the supervised learning , and $p$ is the input dimension. For example, for classification problems, $g(\boldsymbol{\theta};\mathbf{X})$ directly outputs the response class probabilities; and for regression with Gaussian errors and squared-error loss, $g(\boldsymbol{\theta};\mathbf{X})$ represents the conditional mean $\mathbb{E}[Y|\mathbf{X};\boldsymbol{\theta}]$ of a Gaussian $\mathbb{P}(Y|\mathbf{X};\boldsymbol{\theta})$. As in \citet{Zhang2023}, we assume $g(\boldsymbol{\theta};\mathbf{X})$ is fit to the training data via maximum likelihood estimation (MLE), perhaps with regularization. 
{This parametric assumption implies that the method is applicable to models such as generalized linear models (GLMs) and neural networks (where gradients with respect to weights are well-defined). It is not directly applicable to non-parametric or tree-based methods like Random Forests or XGBoost, which do not admit a differentiable likelihood function with respect to fixed parameters.}

Denote the training data to which $g(\boldsymbol{\theta};\mathbf{X})$ is fit by $\{(\mathbf{x}_i, y_i): i=1,2,\cdots, n\}$, assumed to be an i.i.d. sample drawn from the joint distribution $\mathbb{P}(Y,\mathbf{X})$. For each observation $(\mathbf{x}_i, y_i)$, the (Fisher) score vector is defined as 
$\mathbf{s}(\boldsymbol{\theta} ;\mathbf{x}_i, y_i)=\nabla_{\boldsymbol{\theta}} \log \mathbb{P}(y_i \vert \mathbf{x}_i ; \boldsymbol{\theta})$. 
Under certain regularity conditions, if the assumed parametric model is indeed the true data-generating mechanism with true parameters denoted by  $\boldsymbol{\theta}^{0}$, a fundamental result in statistical theory \citep{bickel2015mathematical}[Proposition 3.4.4] states that the expected value of the score vector, when evaluated at $\boldsymbol{\theta}^{0}$, is equal to zero
\begin{equation}\label{theoretical MLE}
\mathbb{E}_{\boldsymbol{\theta}^{0}}\left[\mathbf{s}\left(\boldsymbol{\theta}^{0} ;\mathbf{X}, Y\right) \middle|\mathbf{X}\right] =\int \mathbf{s}\left(\boldsymbol{\theta}^{0} ;\mathbf{X}, Y=y\right) \mathbb{P}\left(Y=y \middle| \mathbf{X} ; \boldsymbol{\theta}^{0}\right) \mathrm{d} y=\mathbf{0}.
\end{equation}

In the context of machine learning, the notion of concept drift refers to the phenomenon where the underlying relationship between $\mathbf{X}$ and $Y$ evolves over time, which can be characterized as a shift in $\mathbb{P}(Y | \mathbf{X}; \boldsymbol{\theta})$. In the parametric setting of \citet{Zhang2023}, concept drift translates to a change in $\boldsymbol{\theta}$. Under certain identifiability conditions, when the parameters change (to some $\boldsymbol{\theta} \neq \boldsymbol{\theta}^{0})$, the score vector mean $\mathbb{E}_{\theta}[\mathbf{s}(\boldsymbol{\theta}^{0};\mathbf{X}, Y)|\mathbf{X}] = \int \mathbf{s}(\boldsymbol{\theta}^{0} ;\mathbf{X}, Y=y) \mathbb{P}(Y=y \vert \mathbf{X}; \boldsymbol{\theta}) \mathrm{d} y$ will differ from zero. 
In light of this, the approach of \citet{Zhang2023} converts concept drift monitoring to the equivalent problem of monitoring for a change in the mean of the score vector $\mathbf{s}(\hat{\boldsymbol{\theta}};\mathbf{X}, Y)$, where $\hat{\boldsymbol{\theta}}$ denotes the MLE of $\boldsymbol{\theta}^{0}$ obtained by fitting the model $g(\boldsymbol{\theta};\mathbf{X})$ to the training data.
The MLE is given by 
\begin{equation}
\hat{\boldsymbol{\theta}} : =\underset{\boldsymbol{\theta}}{\operatorname{argmax}} \frac{1}{n} \sum_{i=1}^n \log \mathbb{P}\left(y_i \middle|\mathbf{x}_i ; \boldsymbol{\theta}\right),
\end{equation}
in which case
\begin{equation}\label{empirical MLE}
\nabla_{\boldsymbol{\theta}} \frac{1}{n} \sum_{i=1}^n \log \mathbb{P}\left(y_i \middle| \mathbf{x}_i ; \boldsymbol{\theta}\right)|_{\boldsymbol{\theta}=\hat{\boldsymbol{\theta}}}=\frac{1}{n} \sum_{i=1}^n\mathbf{s}\left(\hat{\boldsymbol{\theta}} ;\mathbf{x}_i, y_i\right)=\mathbf{0},
\end{equation}
which is the empirical counterpart of \eqref{theoretical MLE}, since it represents the average score vector over the training data.

The empirical counterpart \eqref{empirical MLE} also provides some intuition for why
the score-based concept drift monitoring approach can remain effective even when
the parametric supervised learning model $g(\boldsymbol{\theta};\mathbf{X})$
is only an approximation to the true predictive relationship
$\mathbb{P}(Y \mid \mathbf{X})$. Suppose the predictive relationship between $Y$ and $\mathbf{X}$ for a new set of data $\{(\mathbf{x}_{n+i}, y_{n+i}): i=1, 2, \cdots, m\}$ differs substantially from what it was over the training data. The value of $\boldsymbol{\theta}$ that minimizes the log-likelihood over the new data will generally differ from the MLE $\hat{\boldsymbol{\theta}}$ over the training data, in which case
\begin{equation}\label{shift}
\nabla_{\boldsymbol{\theta}} \frac{1}{m} \sum_{i=1}^{m} \log \mathbb{P}\left(y_{n+i} \middle| \mathbf{x}_{n+i} ; \boldsymbol{\theta}\right)|_{\boldsymbol{\theta}=\hat{\boldsymbol{\theta}}}
     =\frac{1}{m} \sum_{i=1}^{m} \mathbf{s}\left(\hat{\boldsymbol{\theta}} ;\mathbf{x}_{n+i}, y_{n+i}\right),
\end{equation}
will differ from zero. {This argument does not require the supervised learning model to be correctly specified.
Even under model misspecification, the maximum likelihood estimator converges to a
pseudo-true parameter minimizing the Kullback--Leibler divergence.
When the underlying predictive relationship changes, this pseudo-true parameter
typically shifts, implying that the expected score vector evaluated at the original
training fit is no longer zero.
The same arguments hold if one uses a regularized version of MLE with the score vectors
replaced by the derivative of the regularized log-likelihood, which is common in practice.
This robustness property was previously discussed and supported by the empirical results in \citet{Zhang2023} and subsequent follow-ups, including applications to microstructure image data in \citet{Zhang2021}
(for which the supervised learning model is always only an approximate representation of
the true relationship). It is also demonstrated in our nonlinear oscillator example (Section \ref{Nonlinear Oscillator System}), where a simplified neural network only approximates the complex physical dynamics but still allows changes in the physical parameters to be successfully detected. Theorem \ref{thm:misspec} formalizes this detectability argument under model misspecification.}

This was the basis for the concept drift monitoring approach of \citet{Zhang2023}, who used a standard multivariate EWMA (MEWMA) to monitor for changes in the mean of the score vector as the new data are collected. To determine the CL for the MEWMA, they divide the training data $\mathcal{D} = \{(\mathbf{x}_i, y_i): i=1,2,\cdots, n\}$ into two subsets: $\mathcal{D}_1 := \{(\mathbf{x}_i,y_i): i=1, 2, \cdots, n_1\}$ and $\mathcal{D}_2 := \{(\mathbf{x}_i,y_i): i=n_1+1, n_1+2, \cdots, n\}$. The first subset, $\mathcal{D}_1$, is used to fit the supervised learning model $g(\hat{\boldsymbol{\theta}};\mathbf{x})$, producing the MLE $\hat{\boldsymbol{\theta}}$. 
They then compute the score vectors $\mathbf{s}(\hat{\boldsymbol{\theta}}; \mathbf{x}_i,y_i)$ over the second subset, $\mathcal{D}_2$, and apply an MEWMA to these score vectors to empirically compute the CL. Specifically, the CL is taken to be the $1-\alpha$ ($\alpha$ is the desired false-alarm rate) sample quantile of the Hotelling $T^2$ statistics, $\{(\mathbf{z}_i-\overline{\mathbf{s}})^\top \widehat{\boldsymbol{\Sigma}}^{-1}(\mathbf{z}_i-\overline{\mathbf{s}}): i=n_1+1, n_1+2, \cdots, n\}$, where $\overline{\mathbf{s}} = \sum_{i=n_1+1}^{n} \mathbf{s}(\hat{\boldsymbol{\theta}}; \mathbf{x}_i,y_i)/(n-n_1)$ and $\widehat{\boldsymbol{\Sigma}} = \sum_{i=1}^{n_1} (\mathbf{s}(\hat{\boldsymbol{\theta}}; \mathbf{x}_i,y_i) - \overline{\mathbf{s}})(\mathbf{s}(\hat{\boldsymbol{\theta}}; \mathbf{x}_i,y_i) - \overline{\mathbf{s}})^\top/n_1$ are the mean vector and covariance matrix of the score vectors for the second subset, and the MEWMA $\mathbf{z}_i$ is defined recursively for $i = n_1+1, n_1+2, \cdots$ via $\mathbf{z}_i=\lambda \mathbf{s}_i+(1-\lambda)\mathbf{z}_{i-1}$, where $\lambda$ is the MEWMA smoothing parameter.

The method of \citet{Zhang2023} was primarily intended for situations in which $n$ is very large, since the size of $\mathcal{D}_2$ must be quite large to accurately determine the $T^2$ CL using the above procedure. Our nested bootstrap procedure (described in Section \ref{Bias Detection Algorithms and Derivations}) to compute the $T^2$ CL results in much more accurate false-alarm rate control. This is especially true when $n$ is not sufficiently large and/or the desired false-alarm rate is small, because the method of \citet{Zhang2023} requires a very large $\mathcal{D}_2$ for small false-alarm rates (FAR) (e.g., for a desired false-alarm rate of 0.001, the size of $\mathcal{D}_2$ must be much larger than 1,000). Moreover, our procedure allows the CL to be computed from the same data to which the supervised learning model is fit. This removes the need to divide the training sample into two subsets and allows the
entire sample $\mathcal{D}$ to be used for model fitting, which results in a more accurate model.

{Practitioners often encounter scenarios where concept drift affects the relationship between $Y$ and only a subset of the feature vector $\mathbf{X}$ (e.g., the example in Section \ref{Nonlinear Oscillator System}, later).
In such cases, the true parameters associated with the unaffected features may remain constant, while those associated with the drifting features change.
Because the score vector $\mathbf{s}(\hat{\boldsymbol{\theta}};\mathbf{x},y)$ contains components for all model parameters, and the MEWMA statistic $T^2$ aggregates deviations across the entire parameter vector space, the method can detect these partial shifts.
The detection power will of course depend on how strongly the drifting subset influences the overall likelihood.}

\section{Nested Bootstrap Procedure for Computing the CL}\label{Bias Detection Algorithms and Derivations}

Suppose the training sample $\mathcal{D}=\{(\mathbf{x}_i, y_i)\}_{i=1}^n$ is drawn i.i.d. from some joint distribution $\mathbb{P}_0(Y,\mathbf{X})$ for which the conditional distribution $\mathbb{P}(Y|\mathbf{X};\boldsymbol{\theta}^{0})$ can be implicitly represented by the parametric family $g_\gamma(\boldsymbol{\theta};\mathbf{x})$ of supervised learning models, {where $\gamma$ denotes fixed hyperparameters such as regularization constants}. Let $\mathbb{E}_0[\cdot]$ denote the expectation operator with respect to $\mathbb{P}_0(Y,\mathbf{X})$. 

We apply the supervised learning model to the training dataset $\mathcal{D}$ and compute the score vectors. For notational simplicity, we denote these score vectors as $\cS=\{\mathbf{s}_i: i=1,2, \cdots, n\}$ instead of using the functional notation $\mathbf{s}(\hat{\boldsymbol{\theta}};\mathbf{x}_i,y_i)$ from previous sections. Recall, each $\mathbf{s}_i$ is the derivative of the component of the model fitting objective function (the negative log-likelihood or a regularized version, with no optimization constraints) associated with observation $(\mathbf{x}_i, y_i)$. Let $\overline{\mathbf{s}} = 1/n \sum_{i=1}^{n}\mathbf{s}_{i}$ and $\widehat{\boldsymbol{\Sigma}}=1/n \sum_{i=1}^{n}
  (\mathbf{s}_{i}-\overline{\mathbf{s}})
  (\mathbf{s}_{i}-\overline{\mathbf{s}})^{\top}$ denote the sample mean vector and covariance matrix of $\cS$, respectively. By construction, $\overline{\mathbf{s}}=0$.

Now we consider a set of new observations $\mathcal{D}^{\text{new}}=\{(\mathbf{x}_{n+i}, y_{n+i}): i=1,2,\cdots\}$ drawn i.i.d. from the same distribution as $\mathcal{D}$, to represent the situation that there is no shift in the predictive distribution. The score vector $\mathbf{s}_{n+i}= \mathbf{s}(\hat{\boldsymbol{\theta}};\mathbf{x}_{n+i}, y_{n+i})$ for each new observation depends on $(\mathbf{x}_{n+i}, y_{n+i}, \hat{\boldsymbol{\theta}})$. Let $\mathcal{S}^{\text{new}} = \{\mathbf{s}_{n+i}: i = 1, 2, \cdots\}$ denote the new score vectors, and note that $\mathcal{S}^{\text{new}}$ depends on $\mathcal{D}$ only via the fitted model parameters $\hat{\boldsymbol{\theta}}$. Although the regularization parameters $\mathbf{\gamma}$ are also estimated from $\mathcal{D}$, we treat them as fixed for tractability and use the same values when fitting all models within the bootstrapping procedure described below. Consequently, conditioned on $\hat{\boldsymbol{\theta}}$, $\mathcal{S}^{\text{new}}$ constitutes an i.i.d. sample with each $\mathbf{s}_{n+i}$, $i=1,2,\cdots$, having some common mean $\mathbf{\mu}(\hat{\boldsymbol{\theta}})$ and covariance matrix $\mathbf{V}(\hat{\boldsymbol{\theta}})$ that are deterministic functions of $\hat{\boldsymbol{\theta}}$, which we denote by
\begin{equation}\label{score vector dist}
\mathbf{s}_{n+i} \vert \hat{\boldsymbol{\theta}} \sim \operatorname{i.i.d.}(\mathbf{\mu}(\hat{\boldsymbol{\theta}}), \mathbf{V}(\hat{\boldsymbol{\theta}})), \quad i=1,2,\cdots.
\end{equation}
Although $\overline{\mathbf{s}}=\mathbf{0}$, it is not the case that $\mathbf{\mu}(\hat{\boldsymbol{\theta}})=\mathbf{0}$, because of estimation error in $\hat{\boldsymbol{\theta}}$ and because $\mathbf{s}_{n+i}$ are computed for new observations independent of the training data $\mathcal{D}$ to which $g_\gamma(\hat{\boldsymbol{\theta}};\mathbf{x})$ is fit. For similar reasons, $\widehat{\boldsymbol{\Sigma}}$ should not be viewed as an estimator of $\mathbf{V}(\hat{\boldsymbol{\theta}})$.

Let $\{\mathbf{z}_{n+i}: i = 1, 2, \cdots\}$ denote the MEWMAs of the score vectors in $\mathcal{S}^{\text{new}}$, defined recursively as
$\mathbf{z}_{n+i} = \lambda \mathbf{s}_{n+i} + (1 - \lambda) \mathbf{z}_{n+i-1}$, where $\lambda \in (0,1)$ is the smoothing parameter and $\mathbf{z}_{n}=\mathbf{0}$. Write $\mathbf{z}_{n+i}
         = \lambda\bigl[\mathbf{s}_{n+i}
           + (1-\lambda)\mathbf{s}_{n+i-1}
           + (1-\lambda)^2\mathbf{s}_{n+i-2}
           + \cdots
           + (1-\lambda)^{i-1}\mathbf{s}_{n+1}\bigr]$ for
$i = 1,2,\cdots$. From \eqref{score vector dist}, conditioned on $\hat{\boldsymbol{\theta}}$, the conditional mean and covariance of $\mathbf{z}_{n+i}$ are
\begin{equation}
\begin{aligned}
\mathbb{E}_0\!\left[\mathbf{z}_{n+i}\,\middle|\,\hat{\boldsymbol{\theta}}\right]
    &= \bigl[1-(1-\lambda)^{i}\bigr]\,
       \boldsymbol{\mu}(\hat{\boldsymbol{\theta}}),
    \qquad
\operatorname{Cov}_0\!\left[\mathbf{z}_{n+i}\,\middle|\,\hat{\boldsymbol{\theta}}\right]
    &= \frac{\lambda}{2-\lambda}\,
       \bigl[1-(1-\lambda)^{2i}\bigr]\,
       \mathbf{V}(\hat{\boldsymbol{\theta}}).
\end{aligned}
\end{equation}
The derivations in the remainder of this section relate the distribution of $\mathbf{z}_{n+i}$ (for each $i = 1, 2, \cdots$) to the distribution of analogous MEWMA ($\mathbf{z}_i^{b,j}$ from Step IV (b) of Algorithm \ref{A3}) on the bootstrapped score vectors, in order to determine the CL for the $T^2$ chart on $\{\mathbf{z}_{n+i}: i = 1, 2, \cdots\}$ as a function of $i$. Algorithm \ref{A3} provides an overview of our nested bootstrapping procedure to compute the CL. 
 For each outer bootstrap replicate $b$ $(=1,2,\cdots, B_O)$, let $\mathcal{D}^b=\{(\mathbf{x}_i^b, y_i^b): i=1,2,\cdots,n\}$ denote the bootstrap sample of size $n$ from $\mathcal{D}$, and let $\mathcal{D}_{OOB}^b=\{\bigl(\mathbf{x}_{\mathrm{OOB},i}^{\,b},\; y_{\mathrm{OOB},i}^{\,b}\bigr): i\in OOB^b\}$ denote the corresponding out-of-bag (OOB) observations, where $OOB^b$ represents the indices of observations in $\mathcal{D}$ that are not selected in bootstrap sample $b$. The score vectors computed from these samples are denoted as $\mathcal{S}^b=\{\mathbf{s}_i^b: i=1,2,\cdots,n\}$ and $\mathcal{S}_{OOB}^b=\{\mathbf{s}_{OOB,i}^b: i\in OOB^b\}$ respectively, where $\mathbf{s}_{i}^{\,b}
  = \mathbf{s}(\hat{\boldsymbol{\theta}}^{\,b};
                    \mathbf{x}_{i}^{\,b}, y_{i}^{\,b})$, $\mathbf{s}_{\mathrm{OOB},i}^{\,b}
  = \mathbf{s}(\hat{\boldsymbol{\theta}}^{\,b};
                    \mathbf{x}_{\mathrm{OOB},i}^{\,b},
                     y_{\mathrm{OOB},i}^{\,b})$, and $\hat{\boldsymbol{\theta}}^{\,b}$ are the parameters of the model fit to $\mathcal{D}^{\,b}$ in Step 3-II of Algorithm \ref{A3}. 
Let $\overline{\mathbf{s}}^{\,b}
   = 1/n\sum_{i=1}^{n}\mathbf{s}_{i}^{\,b}$ and $\widehat{\boldsymbol{\Sigma}}^{\,b}
   = 1/n\sum_{i=1}^{n}
     (\mathbf{s}_{i}^{\,b}-\overline{\mathbf{s}}^{\,b})
     (\mathbf{s}_{i}^{\,b}-\overline{\mathbf{s}}^{\,b})^{\top}$ denote the sample mean vector and covariance matrix of $\mathcal{S}^b$.
For each inner bootstrap replicate $j$ $(=1,2,\cdots, B_I)$ within outer replicate $b$, we denote the bootstrap sample of score vectors drawn (with replacement) from $\mathcal{S}_{OOB}^b$ as $\{\mathbf{s}_i^{b,j}: i=1,2,\cdots\}$ and their corresponding MEWMA as $\mathbf{z}_i^{b,j}$ from Step IV(b) of Algorithm \ref{A3}.

\begin{algorithm}[H]\label{A3}
\caption{Nested Bootstrap Algorithm for Control Chart Setup}
\KwInput{the full training sample $\mathcal{D}$; {desired false-alarm rate $\alpha$}; and the MEWMA parameter $\lambda$.}
\KwResult{the upper CL $CL_i$ for $i=1,2,3,\cdots$.}

1) Fit and tune a supervised learning model $g_\gamma(\hat{\boldsymbol{\theta}};\mathbf{x})$ to $\mathcal{D}$ using cross validation (CV) to select hyperparameters $\gamma$.

2) Apply the model to compute the score vectors $\mathcal{S}$, and their sample mean $\overline{\mathbf{s}}$ and covariance matrix $\widehat{\boldsymbol{\Sigma}}$.  

3) Using the following nested bootstrapping procedure, determine CL $\{CL_i: i=1,2,3,\cdots\}$ for the $T^2$ statistic.

For $b=1,2,\cdots,B_O$ (outer bootstrap loop):
\begin{enumerate}[label=\Roman*.]
    \item Draw a bootstrap sample $\mathcal{D}^b$ of size $n$ from $\mathcal{D}$, and identify the OOB observations $\mathcal{D}_{OOB}^b$.

    \item Using the same tuning parameters $\gamma$ from Step 1, fit a new model $g_\gamma(\hat{\boldsymbol{\theta}}^{b};\mathbf{x})$ to $\mathcal{D}^b$.

    \item Compute the score vectors $\mathcal{S}^b$ and $\mathcal{S}_{OOB}^b$, and the sample mean $\overline{\mathbf{s}}^b$ and covariance matrix $\widehat{\boldsymbol{\Sigma}}^b$ of $\mathcal{S}^b$.  

    \item For $j=1,2,\cdots,B_I$ (inner bootstrap loop):
    \begin{enumerate}[label=(\alph*)]
        \item Draw a bootstrap sample of score vectors $\{\mathbf{s}_i^{b,j}: i=1,2,3,\cdots\}$ from $\mathcal{S}_{\mathrm{OOB}}^b$. Initialize $\mathbf{z}_0^{b,j}=\mathbf{0}$.
        \item For $i=1,2,3,\cdots$, compute the MEWMA
    $\mathbf{z}_i^{b,j}=\lambda\mathbf{s}_i^{b,j}+(1-\lambda)\mathbf{z}_{i-1}^{b,j}$.
        \item  {For $i=1,2,3,\cdots$, compute the variance correction factor $k(\lambda,i,n)$ as in \eqref{eq:k-def-short}, form $\tilde{\mathbf z}_i^{\,b,j}=k(\lambda,i,n)^{-1/2}\mathbf z_i^{\,b,j}$, and compute the corresponding bootstrap statistic $T_i^{b,j}$ as in \eqref{sampling T statistics2}.}
    \end{enumerate}
\end{enumerate}
For each $i=1,2,3,\cdots$, set $CL_i$ to be the upper $\alpha$ quantile of $\{T_i^{b,j}: b=1,2,\cdots,B_O;\; j=1,2,\cdots,B_I\}$.
\end{algorithm}

{It is worth noting control charts are often designed to give a desired in-control Average Run Length (ARL), as opposed to false alarm rate. However, calculating ARLs via Monte Carlo simulation for our approach is computationally prohibitive due to the nested bootstrap required at each time step.
Therefore, Algorithm \ref{A3} is designed to control the pointwise false-alarm rate $\alpha$ at each observation $i$, which we validate empirically in Section \ref{Numerical Illustrations}.}

{
Within the nested bootstrap procedure, each outer bootstrap sample $\mathcal{D}^b$ in Step~3-I assumes the role of the original training sample $\mathcal{D}$ and accounts for Phase~I (model-estimation)  uncertainty arising from fitting the supervised learning model, i.e., variability in $\hat{\boldsymbol{\theta}}$. The corresponding out-of-bag (OOB) sample $\mathcal{D}_{\mathrm{OOB}}^b$ plays the role of the future data stream $\mathcal{D}^{\mathrm{new}}$ and captures Phase~II (prospective monitoring) variability in incoming observations. As a result, Algorithm~\ref{A3} is designed to control the marginal pointwise false-alarm rate at each monitoring time, explicitly accounting for both Phase~I uncertainty (via the outer bootstrap loop) and Phase~II data variability (via the inner bootstrap loop). In contrast, a single (non-nested) bootstrap procedure would account only for Phase~II variability and therefore substantially underestimate the control limit.}

{Let $n^b$ denote the number of unique observations in $\mathcal{D}^b$, in which case $\lvert \mathcal{D}_{\mathrm{OOB}}^b \rvert = n - n^b$. We employ standard bootstrap sampling with replacement. Consequently, each bootstrap sample $\mathcal{D}^b$ contains approximately the fraction $1 - (1-1/n)^n \approx 0.632$ fraction of unique observations from $\mathcal{D}$ \citep{Efron1983}.
The remaining $\approx 0.368n$ observations form the out-of-bag (OOB) sample $\mathcal{D}_{OOB}^b$, which serves as the population from which independent future data are drawn, with replacement.
For notational simplicity, in the derivations of Section \ref{Covariance Inflation of Naive Bootstrap MEWM and Rescaling for Correct CL Computation}, we assume each $n^b = 0.632 n$.}

\subsection{Covariance Inflation of Naive Bootstrap MEWM and Rescaling for Correct CL Computation}\label{Covariance Inflation of Naive Bootstrap MEWM and Rescaling for Correct CL Computation}
 {For each $i = 1,2,\cdots$, one might consider using the empirical distribution of a naïve version $\{(\mathbf{z}_{i}^{\,b,j}-\overline{\mathbf{s}}^{\,b})^{\top}
  (\widehat{\boldsymbol{\Sigma}}^{\,b})^{-1}
  (\mathbf{z}_{i}^{\,b,j}-\overline{\mathbf{s}}^{\,b})
  : b = 1, 2, \cdots, B_O; j = 1, 2, \cdots, B_I\}$ of the bootstrapped $T^2$ statistics to approximate the distribution of the $T^2$ statistic $T_{n+i} = (\mathbf{z}_{n+i}-\overline{\mathbf{s}})^{\top} \hat{\Sigma}^{-1} (\mathbf{z}_{n+i}-\overline{\mathbf{s}})$ for $\mathcal{S}^{\text{new}}$ and compute its CL. However, this naïve bootstrap distribution can be severely biased due to the finite sample size of $\mathcal{S}_{OOB}^b$, which we state formally in Theorem \ref{thm:inflation-k}.}
  
{To ensure the validity of the proposed control limits, we require specific regularity assumptions and intermediate results regarding the moments of the MEWMA statistics. The detailed theoretical framework, including formal assumptions, intermediate lemmas, and their proofs, is provided in Appendix \ref{app}.}

\begin{theorem}[\textbf{Covariance inflation factor}]
\label{thm:inflation-k}
{Suppose Assumptions~\ref{ass:model} , \ref{bootstrap assumption}, and \ref{ass:mewma} (see Appendix \ref{app}) hold,
and the functions
$\boldsymbol{\mu}(\boldsymbol{\theta})$ and
$\mathbf{V}(\boldsymbol{\theta})$ are differentiable in a neighborhood
of $\boldsymbol{\theta}_0$.
Then, for each $i \ge 1$, the unconditional covariance of the
bootstrap MEWMA statistic $\mathbf{z}^{b,j}_i$ is inflated relative to
that of the true MEWMA statistic $\mathbf{z}_{i}$ by the scalar factor}
\begin{equation}\label{eq:k-def-short}
{k(\lambda,i,n)
 =
\frac{\displaystyle
        \frac{\lambda}{2-\lambda}\bigl[1-(1-\lambda)^{2i}\bigr]
      + \frac{3.72}{n}\bigl[1-(1-\lambda)^i\bigr]^2}
     {\displaystyle
        \frac{\lambda}{2-\lambda}\bigl[1-(1-\lambda)^{2i}\bigr]
      + \frac{1}{n}\bigl[1-(1-\lambda)^i\bigr]^2}},
\end{equation}
{in the sense that}
\begin{equation}\label{new_es}
{\operatorname{Cov}_0\!\,\bigl[\mathbf{z}^{b,j}_i\bigr]
   \cong k(\lambda,i,n)\,   \operatorname{Cov}_0\!\,\bigl[\mathbf{z}_{i}\bigr].}
\end{equation}
\end{theorem}

{
Theorem~\ref{thm:inflation-k} shows that the inner-loop bootstrap MEWMA statistic
$\mathbf z_i^{b,j}$ is a variance-inflated version of the true in-control MEWMA statistic
$\mathbf z_{n+i}$, with an explicit inflation factor $k(\lambda,i,n)$ arising from the
finite size of the out-of-bag sample.
We therefore apply a variance correction to the bootstrap MEWMA statistic before
forming the corresponding Hotelling-type quadratic form.}

{
Specifically, in Algorithm~\ref{A3}, we construct the bootstrap monitoring statistics
\begin{equation}\label{sampling T statistics2}
\tilde{\mathbf z}_i^{\,b,j}
:= k(\lambda,i,n)^{-1/2}\mathbf z_i^{\,b,j},
\qquad
T_i^{b,j}
=
\bigl(\tilde{\mathbf z}_i^{\,b,j}-\bar{\mathbf s}^{\,b}\bigr)^\top
(\hat{\boldsymbol\Sigma}^{\,b})^{-1}
\bigl(\tilde{\mathbf z}_i^{\,b,j}-\bar{\mathbf s}^{\,b}\bigr).
\end{equation}
and define the control limit $\mathrm{CL}_i$ as the empirical $(1-\alpha)$ quantile of
$\{T_i^{b,j}: b=1,\dots,B_O,\; j=1,\dots,B_I\}$.
The variance correction is applied {only} to the bootstrap statistics
$\mathbf z_i^{b,j}$ to compensate for the covariance inflation identified in
Theorem~\ref{thm:inflation-k}; no such correction is applied to the observed MEWMA
statistic $\mathbf z_{n+i}$.}

\begin{remark}\label{remark 3.3}
{
    Equation \eqref{new_es} implies that $\operatorname{Cov}_0[\mathbf{z}_{i}^{\,b,j}]
   = \operatorname{Cov}_0[\mathbf{z}_{i}] + [1 - (1 - \lambda)^{i}]^2\mathbf{I}(\boldsymbol{\theta}^{0})/{0.368n}$, where, from equation ~\eqref{eq:boot-zn-conditional-moments}, the term
$\mathbf{I}(\boldsymbol{\theta}^{0})/0.368n$ ($\mathbf{I}(\boldsymbol{\theta}^{0})$ is the Fisher information matrix) in the difference is precisely
$\mathbb{E}_0[\operatorname{Cov}_0[\overline{\mathbf{s}}_{\mathrm{OOB}}^{\,b}
      \mid\hat{\boldsymbol{\theta}}^{\,b}]]$.
Thus, this difference accounts for the variability in the mean
$\overline{\mathbf{s}}_{\mathrm{OOB}}^{\,b}$ of the ``population''
$\mathcal{S}_{\mathrm{OOB}}^{\,b}$ (whose cardinality is $0.368\,n$) from which
the inner bootstrap samples are drawn in Algorithm \ref{A3}.
This constitutes a nuanced instantiation of the $0.632$ bootstrap rule that is crucial for obtaining a correct CL.
When this correction is omitted (in which case the scaling factor $k(\lambda, i,n)$ in \eqref{sampling T statistics2} is replaced by $1$), we have observed empirically that the bootstrapped CL is typically far too large and the detection power is unnecessarily compromised.}
\end{remark}

{Monitoring is then carried out by comparing
\begin{equation}\label{new T statistics2}
T_{n+i}
=
(\mathbf z_{n+i}-\bar{\mathbf s})^\top
\hat{\boldsymbol\Sigma}^{-1}
(\mathbf z_{n+i}-\bar{\mathbf s})
\end{equation}
to $\mathrm{CL}_i$.
After the control limits are obtained, monitoring proceeds in the usual way: for each
new incoming observation $(\mathbf X_{n+i},Y_{n+i})$, compute its score vector, update
the MEWMA statistic $\mathbf z_{n+i}$, and compare the corresponding $T_{n+i}$ value to
$\mathrm{CL}_i$. A signal is issued at the first time $i$ for which
$T_{n+i} > \mathrm{CL}_i$.}

\begin{remark}[Stabilization of the control limit]
{
Because the MEWMA statistic is initialized at zero, the bootstrap control limits
$\mathrm{CL}_i$ produced by Algorithm~\ref{A3} are time-varying during the early 
stages of monitoring.
However, under standard regularity conditions and in the absence of concept drift,
the bootstrap conditional distributions $F_i(\cdot\mid\mathcal{D})$ converge to a 
limiting distribution $F_\infty(\cdot\mid\mathcal{D})$ as $i\to\infty$, which implies 
that the control limit $\mathrm{CL}_i$ converges to a finite constant.
As a result, after a moderate burn-in period, the control limit is effectively
time-invariant, and recomputation at every time step is unnecessary.
A formal statement and proof of this stabilization result are provided in
Appendix~\ref{app2}.}
\end{remark}

\begin{remark}[Detectability under model misspecification]
{
The calibration results in this section rely on regularity conditions for likelihood-based
estimation, but effective drift detection does not require correct model specification.
Even when the supervised learning model is misspecified, changes in the underlying
predictive relationship typically induce a shift in the pseudo-true parameter that
minimizes Kullback--Leibler divergence.
Such shifts result in a nonzero mean in the monitored score vector, which in turn produces
a detectable signal in the MEWMA statistic.
A formal detectability result under model misspecification is given in Appendix~\ref{app25}.}
\end{remark}

\subsection{Practical Considerations and Computational Aspects}
Several implementation aspects of the algorithm warrant further discussion. When fitting models to bootstrap samples in Step 3-II, it is important to use the same tuning parameters $\gamma$ that were selected in Step 1 rather than performing new CV, as this maintains consistency in the model structure across replicates. If the covariance matrices $\widehat{\boldsymbol{\Sigma}}$ in Step 2 and $\widehat{\boldsymbol{\Sigma}}^b$ in Step 3-III are poorly conditioned (e.g., when $\hat{\boldsymbol{\theta}}^{\,b}$ is high-dimensional and/or has highly correlated components), then $\epsilon \mathbf{I}$ for some small scalar $\epsilon$ should be added to them before they are inverted in \eqref{sampling T statistics2} and \eqref{new T statistics2}. If this is done, it should be done consistently in Steps 2 and 3-III using the same value of $\epsilon$.
Note that $\widehat{\boldsymbol{\Sigma}}^b$ will typically be more poorly conditioned than $\widehat{\boldsymbol{\Sigma}}$, since there are fewer distinct score vectors in $\mathbf{S}_{\mathcal{D}}^b$ than in $\mathbf{S}_{\mathcal{D}}$. Thus, whether $\widehat{\boldsymbol{\Sigma}}$ is replaced by $\widehat{\boldsymbol{\Sigma}}+\epsilon \mathbf{I}$ in Step 2 should depend on how poorly conditioned the $\widehat{\boldsymbol{\Sigma}}^b$ are in step 3-III.

{The choice of monitoring parameters $\alpha$ and $\lambda$ follows standard MEWMA practice. $\alpha$ is determined by the acceptable false-alarm rate, and $\lambda$ balances the trade-off between detecting small, gradual shifts versus large, abrupt shifts \citep{Montgomery2020}.
For the bootstrap parameters, $B_O$ governs the accuracy of estimating the variability due to model fitting, while $B_I$ captures the variability of future data.
Since $B_O$ determines the number of model refits, it represents the primary computational constraint. However, we note that the $B_O$ replicates are independent and can be efficiently performed simultaneously across multiple cores or GPUs. For deep learning models (like the one in Section \ref{Nonlinear Oscillator System}), the refitting process can leverage GPU acceleration to significantly reduce runtime.
For $B_I$, the inner-loop computations are relatively inexpensive.
Accordingly, we recommend that practitioners choose $B_I$ to be conservatively large (e.g., $B_I \ge 200$) and focus computational
resources on maximizing $B_O$, which governs the dominant cost of
model refitting
As demonstrated in Appendix~D (Table~\ref{tab:compute_compare}), the outer bootstrap 
loop is fully parallelizable, reducing the CL setup time to be comparable to or faster 
than that of \citet{Zhang2023} under modest parallelization.}

\begin{remark}
{ In some practical applications, new data may arrive in batches rather than as single observations. In such cases, we recommend randomly ordering the observations within the batch and processing them sequentially through the MEWMA recursion. Because the control limit $CL_i$ is a function of the time index $i$ relative to the start of monitoring (and the training data) but does not depend on the specific values of the new observations, the random ordering of a batch does not alter the threshold sequence.}
\end{remark}

After using Algorithm \ref{A3} to establish the CL, monitoring for concept drift in new observations is straightforward. Given each new observation $(\mathbf{x}_{n+i}, y_{n+i})$, we compute its score vector $\mathbf{s}_{i}$ using the model $g_\gamma(\hat{\boldsymbol{\theta}};\mathbf{x})$, update the MEWMA statistic via $\mathbf{z}_{n+i} = \lambda\mathbf{s}_{n+i} + (1-\lambda)\mathbf{z}_{n+i-1}$,
and compute the $T^2$ monitoring statistics \eqref{new T statistics2}.

Concept drift is detected when $T_{n+i}$ exceeds its corresponding CL $CL_i$. The choice of the MEWMA smoothing parameter $\lambda$ also affects the monitoring sensitivity, and the tradeoff is the same as in any MEWMA monitoring procedure:  Smaller values of $\lambda$ provide more powerful eventual detection of large shifts but can delay the detection of large shifts, whereas larger values provide quicker detection of large shifts but may fail to detect smaller shifts.

\section{Numerical Examples}\label{Numerical Illustrations}

To illustrate our approach and demonstrate its effectiveness at controlling the false-alarm rate, we present two numerical examples. The first is a more transparent example in which the predictive relationship $\mathbb{P}(Y\mid \mathbf{X})$ is a mixture of two simple linear relationships, and the second involves a more complex nonlinear predictive relationship.

\subsection{Mixed Linear Population} \label{mlp41}

The data-generating process for this example involves the two linear models
\begin{equation}\label{curve 1}
y_i = 16x_i + 5 + \varepsilon_i,
\qquad 
\varepsilon_i \sim \mathcal{NID}(0,16),
\end{equation}
and
\begin{equation}\label{curve 2}
y_i = 12x_i + 3 + \varepsilon_i, 
\qquad 
\varepsilon_i \sim \mathcal{NID}(0,16),
\end{equation}
where predictor values $x_i$ are uniformly sampled from $[-\sqrt{3},\sqrt{3}]$.
The training data of size $n = 2,000$ are generated exclusively from
model~\eqref{curve 1}, yielding an pre-shift sample governed by a single linear
relationship.
The future data to be monitored comprise 1,000 observations: the first
200 points (pre-shift) follow the same single linear model~\eqref{curve 1},
while the remaining 800 points (post-shift) are generated from a mixture
that draws $y_i$ from either \eqref{curve 1} or \eqref{curve 2} with equal
probability~0.5, representing a shift in $\mathbb{P}(Y|\mathbf{X})$.

We fit a linear predictive model $g_\gamma(\hat{\boldsymbol{\theta}};x)=\hat{\theta}_0+\hat{\theta}_1x$
via ridge regression with $L_2$ regularization parameter~$\gamma=0.1$ (i.e. with the loss function
$\sum_{i=1}^{n}[(y_i-\mathbf{x}_{i}^{\top}\hat{\boldsymbol{\theta}})^2
      +\gamma\|\hat{\boldsymbol{\theta}}\|^{2}/n],$
for which the penalized log-likelihood for observation~$i$ is
\[
\ell_{\text {pen }}(\hat{\boldsymbol{\theta}};\mathbf{x}_i, y_i) = \log\mathbb{P}\bigl(y_i\mid\mathbf{x}_i;\hat{\boldsymbol{\theta}}\bigr)-\frac{\gamma}{2 n}\|\hat{\boldsymbol{\theta}}\|^2
  = -\frac{1}{2}\bigl(y_i-\mathbf{x}_{i}^{\top}\hat{\boldsymbol{\theta}}\bigr)^{2}
    -\frac12\log\bigl(2\pi\sigma^{2}\bigr)-\frac{\gamma}{2 n}\|\hat{\boldsymbol{\theta}}\|^2.
\]
The score vectors are therefore
\begin{equation}
\mathbf{s}(\hat{\boldsymbol{\theta}}; \mathbf{x}_i, y_i) 
   = \bigl(y_i - \mathbf{x}_{i}^{\top}\hat{\boldsymbol{\theta}}\bigr)\mathbf{x}_{i}
     - \frac{\gamma}{n}\,\hat{\boldsymbol{\theta}}.
\label{eq:score}
\end{equation}
We used MEWMA parameter $\lambda = 0.01$, which we chose via 5-fold CV, and desired pointwise false-alarm
probability $\alpha = 0.001$. In Algorithm \ref{A3}, we used $B_O=100$ and $B_I=200$.

\begin{figure}[!htb]
\centering
\includegraphics[width=4in]{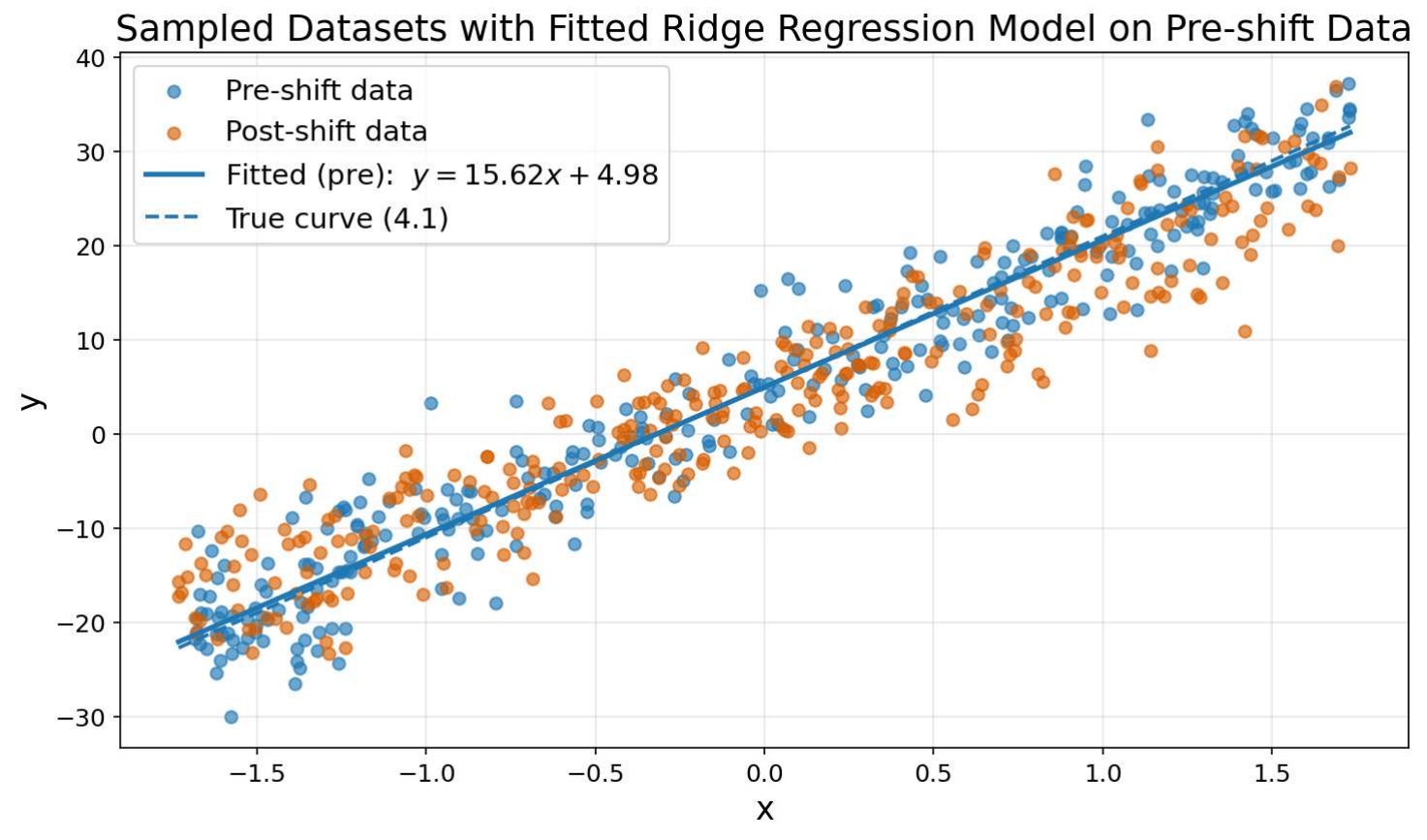}
\caption{Visualization of predictive relationships for the linear example
before and after the shift.
The blue points are a scatter plot of $y$ vs.\ $x$ for the pre-shift
single linear model~\eqref{curve 1}, and the orange points correspond
to the post-shift mixture model ~\eqref{curve 1} and~\eqref{curve 2}.
The shift from a single component to a mixture alters
$\mathbb{P}(Y|\mathbf{X})$, representing a structural shift in the predictive
relationship that is relatively small but can still be detected by our approach.}
\label{fig:dataset}
\end{figure}

\begin{figure}[!htb]
\centering
\includegraphics[width=0.48\textwidth]{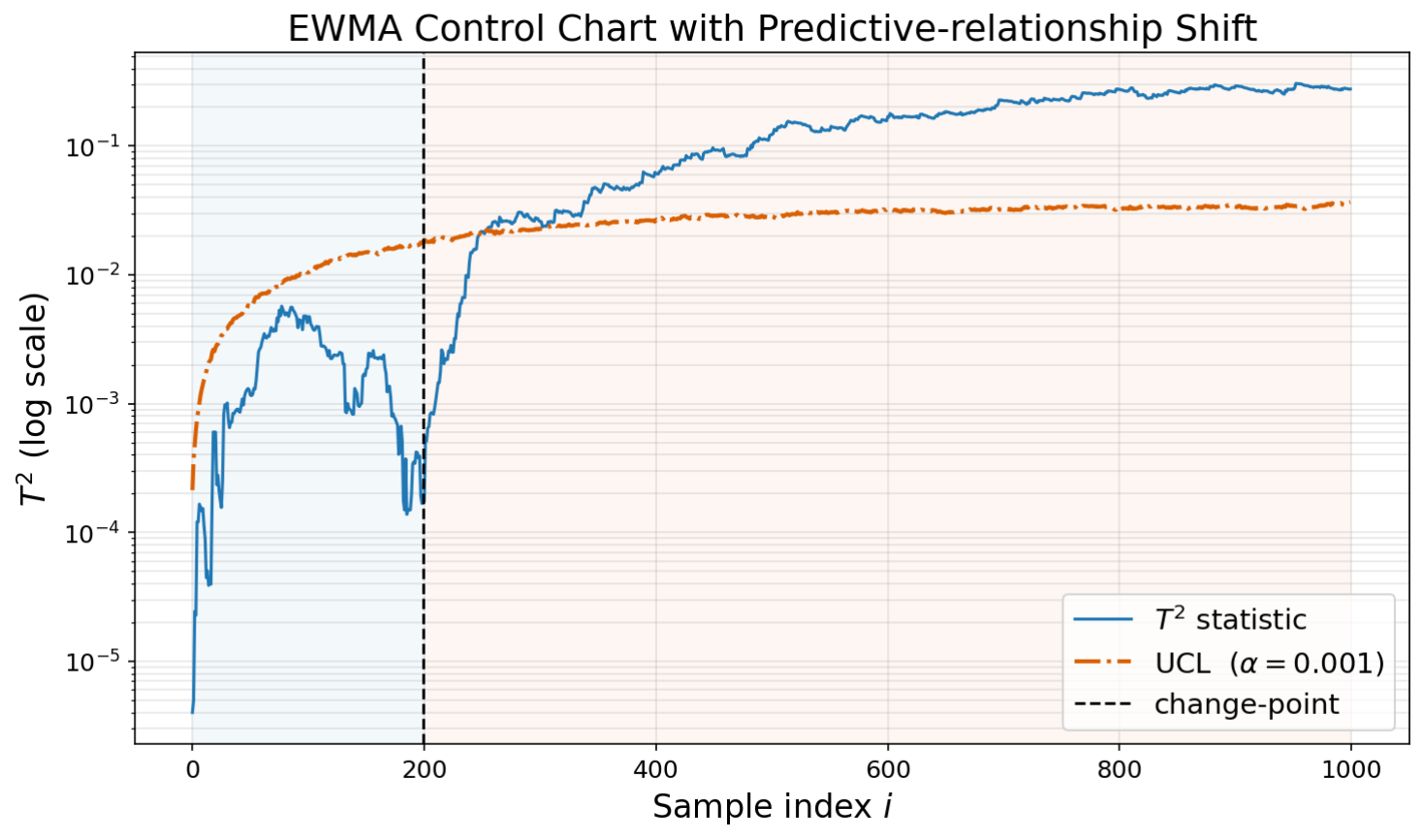}%
\hfill
\includegraphics[width=0.48\textwidth]{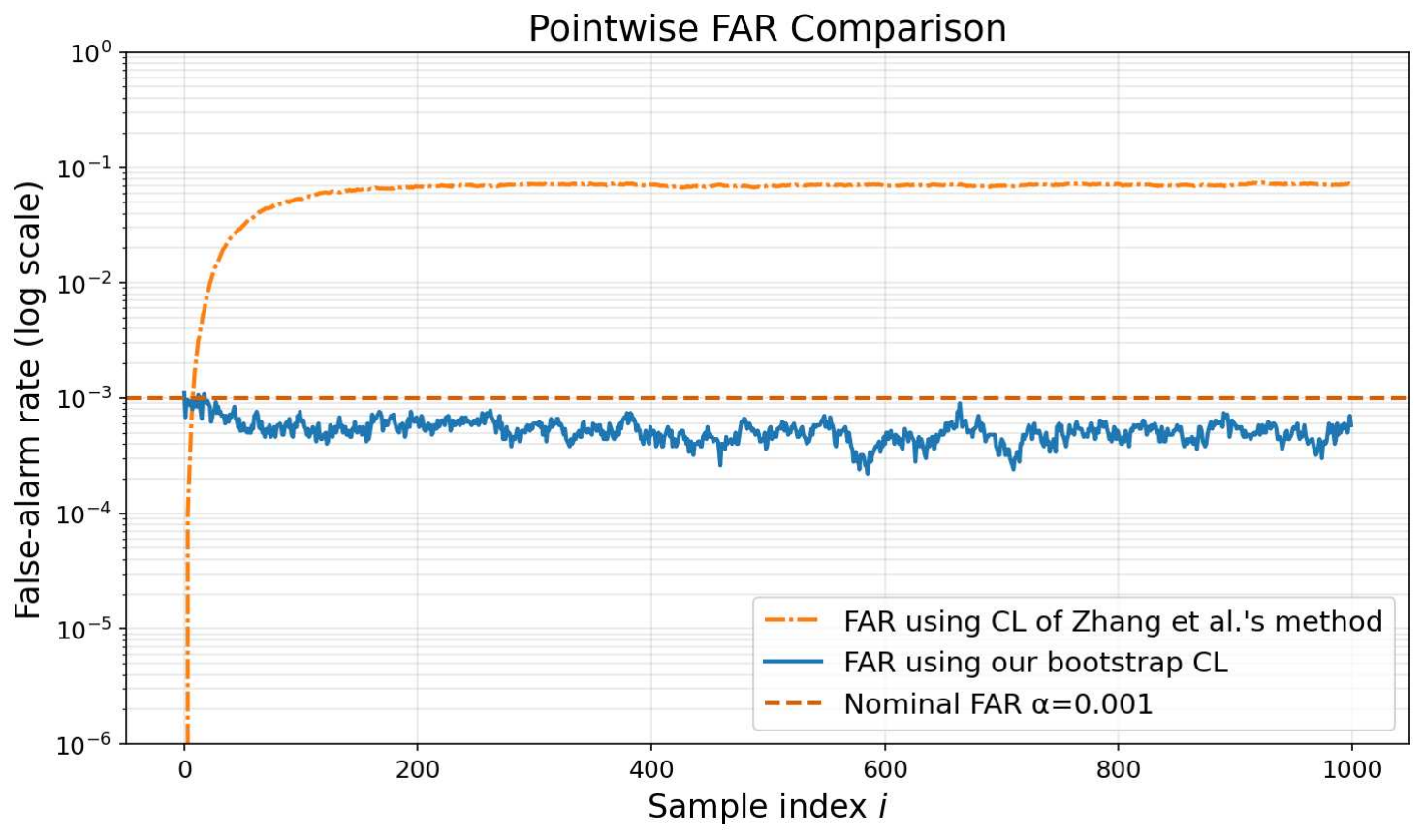}

\vspace{0.5em}
\makebox[0.48\textwidth][c]{(a)}\hfill\makebox[0.48\textwidth][c]{(b)}

\caption{(a) Typical monitoring results using our bootstrap CL approach for the linear mixture example showing the detection of concept drift at new observation number 258 (the shift first occurred at observation number 201). The $T^2$ statistics ($T_i$ from Eq.~\eqref{new T statistics2}, represented by the blue line) remain below the control limit (dashed orange line) prior to the shift at observation 201, and sharply exceed it shortly after the shift. (b) Comparison of the pointwise false-alarm rate for our bootstrap CL (blue curve) versus the CL of \citet{Zhang2023} (orange curve) averaged over 50 database replicates under the pre-shift predictive relationship~\eqref{curve 1}. Our bootstrap CL much more accurately maintains the empirical false-alarm rate close to the desired $\alpha = 0.001$.}
\label{fig:linear_two_panel}
\end{figure}

In Figure~\ref{fig:dataset}, we visualize the transition from the pre-shift predictive relationship \eqref{curve 1} to the post-shift relationship that is a mixture of \eqref{curve 1} and \eqref{curve 2}. 
Although the shift appears quite small in Figure~\ref{fig:dataset}, Figure~\ref{fig:linear_two_panel}(a) demonstrates that it is still detectable. 
Figure~\ref{fig:linear_two_panel}(a) shows a typical evolution of our $T^2$ monitoring
statistic over 1,000 observations, with the first 200 following the pre-shift distribution and the last 800 the post-shift distribution. The $T^2$ statistic remains below the CL prior to the shift and then begins to increase immediately following the shift at time 201 until it first exceeds the CL at time 258, demonstrating the method’s ability to detect the shift.
This behaviour aligns with our derivation in
Section~\ref{Bias Detection Algorithms and Derivations}, where changes in
$\mathbb{P}(Y|\mathbf{X})$ were shown to induce non-zero means in the score
vectors.

To better assess the empirical false-alarm control, the blue curve in Figure~\ref{fig:linear_two_panel}(b) shows the pointwise false-alarm rate over the first 1,000 observations when no shift occurs, estimated as the average pointwise false-alarm rate over 50 replicates. On each replicate, a different training data set was generated, the ridge regression model was fit to these data, Algorithm~\ref{A3} was used to compute the CL as a function of time, a new set of 1,000 observations was generated from the same distribution (representing no shift), and then the $T^2$ statistic was computed for each new observation and compared to its CL to determine if there was a signal at that time. Our method consistently maintains the false-alarm rate at close to the desired value $\alpha = 0.001$.
The orange curve shows the corresponding results from the baseline method of \citet{Zhang2023}, which exhibits substantially inflated FAR exceeding 0.06, highlighting the more accurate false-alarm rate provided by bootstrap CL.

The discrepancy arises for two reasons. First, \citet{Zhang2023} used a constant CL, because without bootstrapping there is no clear way to compute a time-varying CL. Second, \citet{Zhang2023} fit their model on only the first half of the training data and compute the constant CL from the second half. The resulting CL is based on a much smaller effective sample size than in our nested bootstrapping procedure and cannot reliably provide correct Type I error control unless the training sample size is quite large. In contrast, our approach provides accurate false-alarm rate control even with the relatively small training sample size of 2,000 in this example.

It should be noted that the intended application scenarios in \citet{Zhang2023} involved much larger training sample sizes and/or larger values for the desired $\alpha$. For $\alpha = 0.01$ (for example), the method of \citet{Zhang2023} provides much more accurate false-alarm control than in this example with $\alpha = 0.001$. With $\alpha = 0.001$, the method of \citet{Zhang2023} generally requires that the size of the training subset used to compute the CL is at least 10,000 to accurately estimate the upper 0.001 quantile of the empirical distribution of the $T^2$ statistic. In this regard, our bootstrap CL approach can be viewed as an extension of \citet{Zhang2023} to smaller training sample sizes and/or larger desired $\alpha$ values.


\subsection{Nonlinear Oscillator System}\label{Nonlinear Oscillator System}
To demonstrate our method's effectiveness in detecting changes in complex nonlinear predictive relationships, we consider a physics-based system that represents a variant of the two-degree-of-freedom nonlinear oscillator with a finite-extensibility coupling spring proposed by \citet{Febbo2013}. The system consists of two masses moving in one dimension, connected by nonlinear
springs. Let $p_1(t)$ and $p_2(t)$ denote the positions at (continuous) time $t$ of
the two masses having masses $m_1$ and $m_2$, respectively, and let
$v_1(t) = \dot{p}_1(t)$ and $v_2(t) = \dot{p}_2(t)$ denote their velocities.
The system evolves according to the following differential equations:
\begin{equation}\label{physics}
\begin{aligned}
\ddot{p}_1(t) &= \frac{1}{m_1}\bigl(-k_1 p_1(t) - c_1 \dot{p}_1(t)
                 + k_3 \phi(p_1(t), p_2(t))\bigr), \\
\ddot{p}_2(t) &= \frac{1}{m_2}\bigl(-k_2 p_2(t) - c_2 \dot{p}_2(t)
                 - k_3 \phi(p_1(t), p_2(t))\bigr),
\end{aligned}
\end{equation}
where $\phi(p_1,p_2) = (p_1 - p_2)/(1 + |p_1 - p_2|)$ represents the nonlinear
coupling force.

{To generate the training data, each observation corresponds to an independently generated oscillator trajectory.
For each observation $i$, the oscillator system in \eqref{physics} is independently initialized with initial conditions $(p_1(0), v_1(0), p_2(0), v_2(0))$ drawn independently as $p_1(0) \sim \mathrm{Uniform}(0.25, 0.75)$, $v_1(0) \sim \mathrm{Uniform}(-0.25, 0.25)$, $p_2(0) \sim \mathrm{Uniform}(-0.75, -0.25)$, $v_2(0) \sim \mathrm{Uniform}(-0.25, 0.25)$, 
and is then evolved according to \eqref{physics} over the time interval $t \in [0,30]$.
The data for the $i$th trajectory are the states sampled at a fixed grid of $T=20$ equally spaced time points $\{t_1,\ldots,t_T\}\subset[0,30]$.
This yields an i.i.d.\ training sample $\{(\mathbf X_i, y_i)\}_{i=1}^{n}$ with $n=3{,}000$ i.i.d.\ trajectories (identically distributed, because the initial conditions are generated randomly from the same distribution).}

{More specifically, the feature vector $\mathbf{X}_i \in \mathbb{R}^{4T}$ encodes the sampled trajectory by concatenating the system state at the grid times:
\[
\mathbf X_i =
\bigl[
p_{1,i}(t_1), v_{1,i}(t_1), p_{2,i}(t_1), v_{2,i}(t_1),\; 
\ldots,\;
p_{1,i}(t_T), v_{1,i}(t_T), p_{2,i}(t_T), v_{2,i}(t_T)
\bigr]^\top.
\]
Thus, temporal dependence is represented internally within each feature vector, while the observations $\{(\mathbf X_i,y_i)\}$ are i.i.d.\ across trajectories.
The response $y_i$ is defined as the total mechanical energy evaluated at a fixed measurement time $t^\star = 12$, given by
\begin{equation}
y_i = \frac{1}{2}\bigl(m_1 v_{1,i}^2(t^\star) + m_2 v_{2,i}^2(t^\star)\bigr)
      + \frac{1}{2}\bigl(k_1 p_{1,i}^2(t^\star) + k_2 p_{2,i}^2(t^\star)\bigr)
      + k_3 \phi\!\left(p_{1,i}(t^\star), p_{2,i}(t^\star)\right)
      + \varepsilon_i,
\end{equation}
where $(p_{1,i}(t^\star),v_{1,i}(t^\star),p_{2,i}(t^\star),v_{2,i}(t^\star))$ is obtained from the same trajectory and $\varepsilon_i \sim \mathcal{NID}(0,\sigma^{2})$ represents measurement noise.
Note that $t^\star$ does not correspond to any of the $T=20$ fixed time points at which the state variable features in $\mathbf{X}_i$ are recorded. Consequently, the response $y_i$ depends most heavily on the features in $\mathbf{X}_i$ corresponding to times closer to $t^\star$, but it also depends implicitly on other features in $\mathbf{X}_i$, since Eq.~\eqref{physics} evolves deterministically and the states at any time can be computed as a function of the initial conditions. This creates a more challenging drift detection scenario in which the dependence of $y_i$ on the features in $\mathbf{X}_i$ is somewhat sparse.
For the training data, we used the system parameters
$m_1 = 1.0$, $m_2 = 2.0$, $k_1 = 1.0$, $k_2 = 2.0$, $k_3 = 1.5$,
$c_1 = 0.1$, and $c_2 = 0.2$.}

{To simulate concept drift, we generate new oscillator trajectories from the same system but with modified physical parameters
$m_1' = 1.1 m_1, m_2' = 1.2 m_2, k_1' = 1.3 k_1$, while holding the remaining parameters fixed.
These changes represent interpretable real-world perturbations, such as increased
mass (e.g., due to material accumulation) or stiffened springs (e.g., due to thermal
effects), and they alter both the system’s natural frequencies and its energy
landscape. As a result, the trajectory-level conditional distribution $\mathbb{P}(Y \mid \mathbf X)$
is systematically shifted, inducing concept drift that our monitoring framework
aims to detect.}

{To model the nonlinear predictive relationship, we fit a feed-forward multilayer perceptron (MLP)
with $L=4$ hidden layers of width $H=32$ ReLU activation functions, followed by a linear output layer.
That is, the network has the form
\[
h_0(\mathbf x)=\mathbf x,\qquad
h_{\ell}(\mathbf x)=\mathrm{ReLU}(W_{\ell}h_{\ell-1}(\mathbf x)+\mathbf b_{\ell})\ \ (\ell=1,\ldots,L),
\qquad
g_\gamma(\boldsymbol\Theta;\mathbf x)=\mathbf w^\top h_L(\mathbf x)+b.
\]
Here, $x$ generically denotes the outputs of each layer, which are inputs to the subsequent layer, $W_{\ell}\in\mathbb{R}^{H\times H}$ and $\mathbf b_{\ell}\in\mathbb{R}^{H}$ denote the
weights and biases of the $\ell$th hidden layer (with $W_{1}\in\mathbb{R}^{H\times 4T}$),
and $(\mathbf w,b)$ are the parameters of the final linear output layer, and $\Theta$ denotes the complete set of weights and biases to be estimated during training.
This architecture has a few thousand trainable parameters in total, of which $H+1=33$
correspond to the final layer.
}

{Following \citet{Zhang2023}, we monitor only the parameters of the final linear output
layer, denoted by $\boldsymbol\theta=(\mathbf w,b)$, while treating the parameters of the
hidden (feature-extraction) layers
$\boldsymbol\Theta_{\mathrm{feat}}=\{(W_\ell,\mathbf b_\ell)\}_{\ell=1}^L$
as fixed during monitoring.
This strategy keeps the dimension of the score vector tractable even for deep architectures,
while still capturing changes in the predictive relationship
$\mathbb{P}(Y \mid \mathbf X)$ that propagate through the network
(see also \citet{Yosinski2014}).
When fitting the model to the training data, all parameters in $\boldsymbol\Theta$ are
estimated; however, for monitoring, inference is conducted with respect to
$\boldsymbol\theta$ only, using the penalized Gaussian log-likelihood
\[
\ell_{\mathrm{pen}}(\boldsymbol\theta; \mathbf X_i, y_i)
= -\frac{1}{2\sigma^2}
  \bigl(y_i - g_\gamma(\boldsymbol\Theta; \mathbf X_i)\bigr)^2
  - \frac{1}{2}\log(2\pi\sigma^2)
  - \frac{\gamma}{2n}\|\boldsymbol\theta\|^2
  - \frac{\gamma}{2n}\|\boldsymbol\Theta_{\mathrm{feat}}\|^2,
\]
with regularization parameter $\gamma = 10^{-1}$.}
{Hyperparameters were selected using a grid search over the regularization 
parameter $\gamma \in \{10^{-2}, 10^{-1}, 1\}$ and learning rate, minimizing the 
5-fold cross-validation error sum of squares on the training data, yielding 
$\gamma = 10^{-1}$. The MEWMA smoothing parameter was set to $\lambda = 0.01$, 
consistent with Section~\ref{mlp41}.}
{For each trajectory-level observation $(\mathbf X_i, y_i)$, the score vector is defined as the
gradient of the penalized log-likelihood with respect to the monitored parameters
$\boldsymbol\theta$, evaluated at the fitted values
$\hat{\boldsymbol\theta}$.
The score vector is computed via automatic differentiation by calling
\texttt{loss.backward()} in PyTorch.
}

{We consider two scenarios with different noise levels.
The low-noise scenario ($\sigma=0.03$) corresponds to a high signal-to-noise regime, in which departures from the in-control predictive relationship are readily detectable.
The high-noise scenario ($\sigma=0.20$) corresponds to substantially increased observation variability, resulting in a lower signal-to-noise ratio and a more challenging detection problem.}

\FloatBarrier
\begin{figure}[!htbp]
\centering
\begin{subfigure}[b]{0.49\textwidth}
  \centering
  \includegraphics[width=\textwidth]{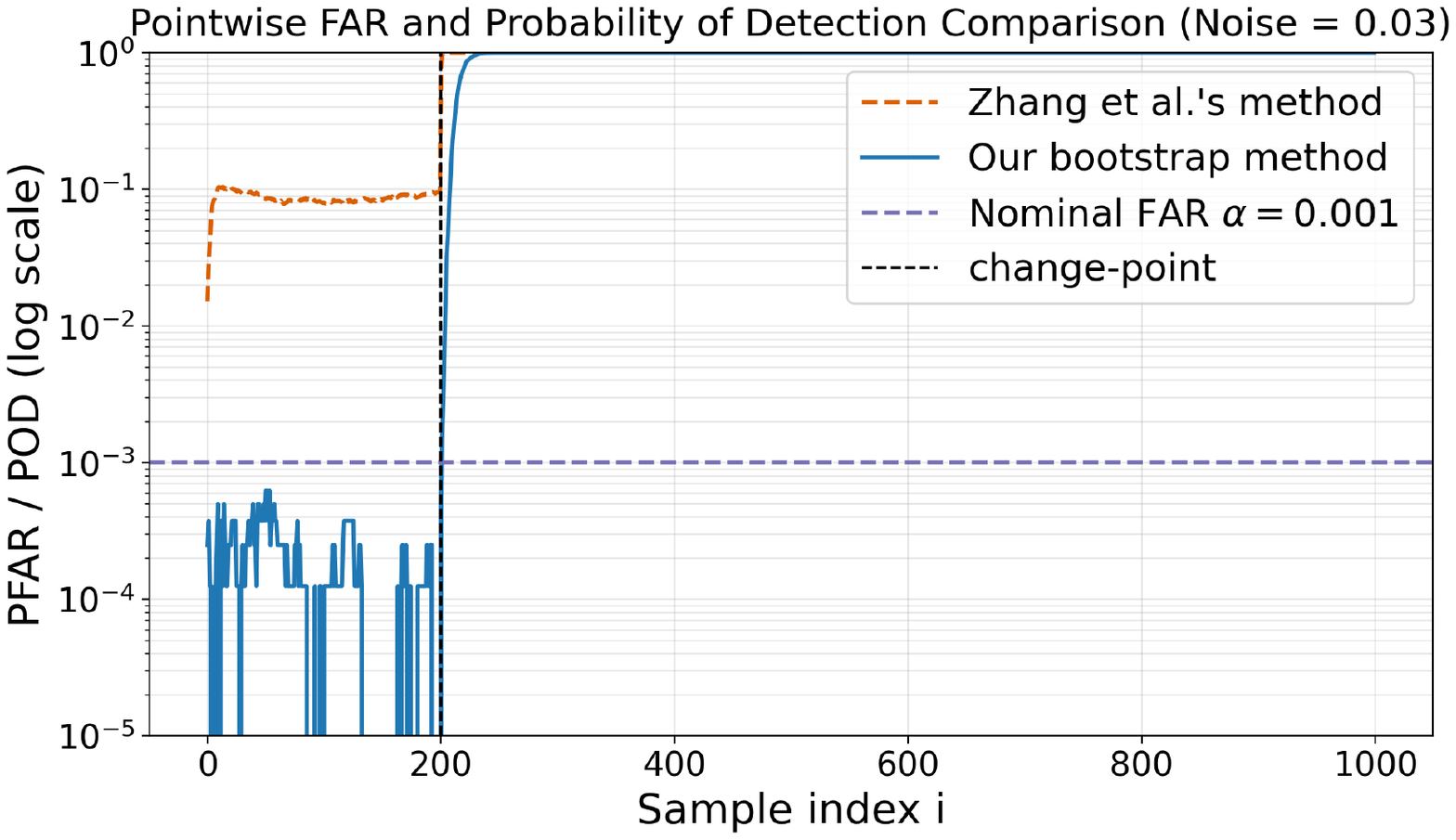}
  \caption{Low noise ($\sigma=0.03$)}
\end{subfigure}
\hfill
\begin{subfigure}[b]{0.49\textwidth}
  \centering
  \includegraphics[width=\textwidth]{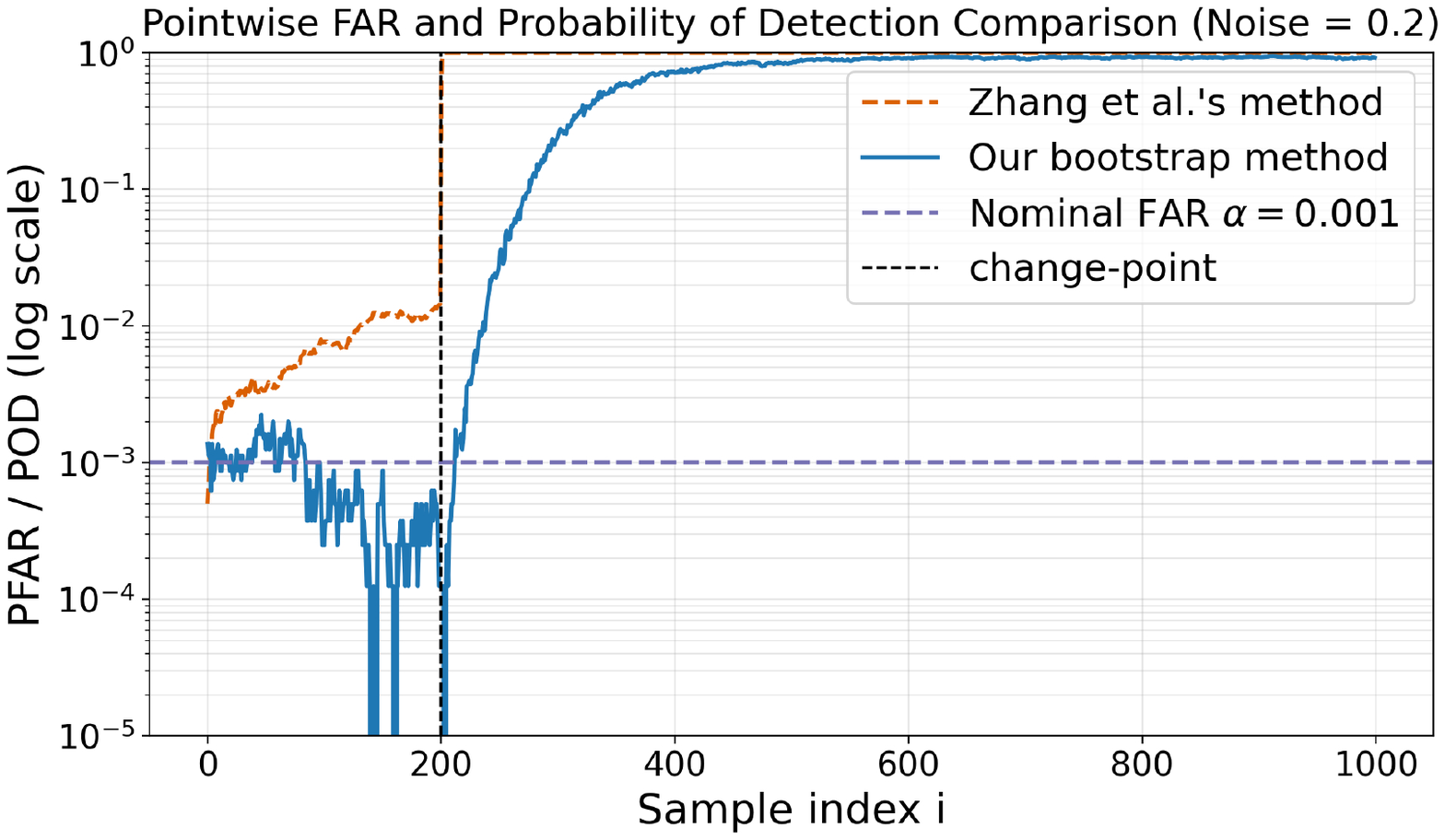}
  \caption{High noise ($\sigma=0.20$)}
\end{subfigure}
\caption{{Empirical pointwise false-alarm rate (PFAR) for the nonlinear oscillator
example under low- and high-noise conditions.
The dashed horizontal line indicates the nominal level $\alpha=0.001$.
Prior to the change point ($i\le 200$), the empirical PFAR remains close to $\alpha=0.001$ in both regimes.
After the change point, the same curves quantify the pointwise probability of detection, which rapidly increases towards one, indicating strong eventual detectability.}}
\label{fig:oscillator_pfar}
\end{figure}

{Figure~\ref{fig:oscillator_pfar} shows the PFAR (before the change was introduced at observation $i=200$) and the pointwise probability of detection (after the change at $i=200$), averaged across 8,000 Monte Carlo replicates. On each Monte Carlo replicate, a different set of training data and future monitoring data were generated as described above for a single replicate. Taken together, these results illustrate that the score-based monitoring approach
remains effective even under substantially increased observation noise, although the average detection delay is longer in the high-noise setting.
Importantly, Figure~\ref{fig:oscillator_pfar} also demonstrates that our bootstrapped CLs maintain the PFAR at close to (or below) the desired $\alpha=0.001$ value. In contrast, the \citet{Zhang2023} control limits resulted in a substantially inflated PFAR, e.g., nearly 100 times larger than $0.001$ in the low-noise setting. Although the \citet{Zhang2023} CL resulted in quicker detection after the change, the detection comparison is not meaningful, since the PFAR was so substantially inflated.}

\section{Conclusions} \label{conclude}
In this paper, we present a nested bootstrapping procedure to compute a time-varying CL for the score-based MEWMA control chart of \citet{Zhang2023}, for detecting changes in the predictive distribution $\mathbb{P}(Y|\mathbf{X})$ represented by a supervised learning model. Our approach provides much more accurate false-alarm rate control than the approach of \citet{Zhang2023}, especially with smaller training sample size $n$ and/or larger desired FAR, and also produces a time-varying CL to maintain accurate false-alarm rate control prior to the MEWMA reaching steady state. Our derivations involve a non-obvious 0.632-type bootstrap correction that is essential for controlling the false-alarm rate. 

{Our approach provides much more accurate false-alarm rate control than the approach of \citet{Zhang2023}, especially with smaller training sample size $n$ and/or strict FAR (e.g., $\alpha=0.001$).
It is important to note that because our method uses the identical monitoring statistic as \citet{Zhang2023}, the theoretical detection power for a given control limit is identical.
The primary advantage of the proposed bootstrap method is the accurate calibration of that limit.
Where the empirical quantile method of \citet{Zhang2023} tends to underestimate the limit in small samples (leading to excessive false alarms) or requires massive holdout sets, our bootstrap correction ensures the nominal Type I error rate is maintained, allowing for valid monitoring even with limited data.
Finally, despite the nested bootstrap structure, the method is computationally practical: 
the outer loop is fully parallelizable, and with modest parallelization the CL setup time 
is comparable to or faster than that of \citet{Zhang2023} (Table~\ref{tab:compute_compare}), 
while providing substantially more accurate false-alarm calibration.}

\section*{Disclosure statement}
The authors have no conflicts of interest to declare.

\bibliography{bibliography.bib}

@article{MorenoTorres2012,
  title={A unifying view on dataset shift in classification},
  author={Jose Garcia Moreno-Torres and Troy Raeder and Roc{\'i}o Ala{\'i}z-Rodr{\'i}guez and N. Chawla and Francisco Herrera},
  journal={Pattern Recognit.},
  year={2012},
  volume={45},
  pages={521-530},
}

@book{bickel2015mathematical,
  title={Mathematical Statistics: Basic Ideas and Selected Topics},
  author={Bickel, Peter J. and Doksum, Kjell A.},
  year={2015},
  publisher={CRC Press},
  address={Boca Raton, FL},
  series={Chapman \& Hall/CRC Texts in Statistical Science},
  volume={1}
}

@article{Zhang2023,
  author = {Kungang Zhang and Anh T. Bui and Daniel W. Apley},
  title = {Concept Drift Monitoring and Diagnostics of Supervised Learning Models via Score Vectors},
  journal = {Technometrics},
  volume = {65},
  number = {2},
  pages = {137--149},
  year = {2023},
  doi = {10.1080/00401706.2022.2124310}
}

@article{Malinovskaya2023,
  author       = {Anna Malinovskaya and Pavlo Mozharovskyi and Philipp Otto},
  title        = {Statistical Process Monitoring of Artificial Neural Networks},
  journal      = {Technometrics},
  volume       = {66},
  number       = {1},
  pages        = {104--117},
  year         = {2024},
  publisher    = {Taylor \& Francis},
  doi          = {10.1080/00401706.2023.2239886},
}

@book{Montgomery2020,
  author    = {Douglas C. Montgomery},
  title     = {Introduction to Statistical Quality Control},
  publisher = {John Wiley \& Sons},
  edition   = {8th},
  year      = {2020},
  isbn      = {978-1119721403}
}

@article{Psarakis2011,
  author = {S. Psarakis},
  title = {The Use of Neural Networks in Statistical Process Control Charts},
  journal = {Quality and Reliability Engineering International},
  volume = {27},
  number = {5},
  pages = {641--650},
  year = {2011}
}

@article{Ross2012,
  author = {G. J. Ross and N. M. Adams and D. K. Tasoulis and D. J. Hand},
  title = {Exponentially Weighted Moving Average Charts for Detecting Concept Drift},
  journal = {Pattern Recognition Letters},
  volume = {33},
  pages = {191--198},
  year = {2012}
}

@inproceedings{Gama2004,
  author = {J. Gama and P. Medas and G. Castillo and P. Rodrigues},
  title = {Learning with Drift Detection},
  booktitle = {Brazilian Symposium on Artificial Intelligence},
  pages = {286--295},
  year = {2004}
}

@article{woodall2004using,
  title={Using control charts to monitor process and product quality profiles},
  author={Woodall, William H and Spitzner, Dan J and Montgomery, Douglas C and Gupta, Shilpa},
  journal={Journal of Quality Technology},
  volume={36},
  number={3},
  pages={309--320},
  year={2004},
  publisher={Taylor \& Francis}
}

@article{Webb2016,
  title={Characterizing concept drift},
  author={Webb, Geoffrey I and Hyde, Roy and Cao, Hong and Nguyen, Hai Long and Petitjean, Francois},
  journal={Data Mining and Knowledge Discovery},
  volume={30},
  number={4},
  pages={964--994},
  year={2016},
  publisher={Springer}
}

@article{Zliobaite2016,
  title={An overview of concept drift applications},
  author={{\v{Z}}liobait{\.e}, Indr{\.e} and Pechenizkiy, Mykola and Gama, Jo{\~a}o},
  journal={Big data analysis: new algorithms for a new society},
  pages={91--114},
  year={2016},
  publisher={Springer}
}

@article{Gama2014,
  title={A survey on concept drift adaptation},
  author={Gama, Jo{\~a}o and {\v{Z}}liobait{\.e}, Indr{\`e} and Bifet, Albert and Pechenizkiy, Mykola and Bouchachia, Abdelhamid},
  journal={IEEE Transactions on Knowledge and Data Engineering},
  volume={27},
  number={3},
  pages={701--720},
  year={2014},
  publisher={IEEE},
  doi={10.1109/TKDE.2014.2341435}
}

@article{Krawczyk2017,
  title={Ensemble learning for data stream analysis: A survey},
  author={Krawczyk, Bartosz and Minku, Leandro L and Gama, Joao and Stefanowski, Jerzy and Wozniak, Michal},
  journal={Information Fusion},
  volume={37},
  pages={132--156},
  year={2017},
  publisher={Elsevier},
  doi={10.1016/j.inffus.2017.02.004}
}

@inproceedings{Baena2006,
  author    = {Manuel Baena{-}Garc\'{\i}a and
               Jos\'{e} del Campo{-}{\'{A}}vila and
               Ra\'{u}l Fidalgo{-}Merino and
               Albert Bifet and
               Ricard Gavald\`{a} and
               Rafael Morales{-}Bueno},
  title     = {Early Drift Detection Method},
  booktitle = {Proceedings of the 4th ECML PKDD International Workshop on
               Knowledge Discovery from Data Streams (KDDS)},
  pages     = {77--86},
  address   = {Berlin, Germany},
  publisher = {Springer},
  year      = {2006}
}

@article{Razak2023,
  author  = {Razak, Abdul M. S. and Nirmala, C. R. and Sreenivasa, B. R. and Lahza, Husam and Lahza, Hassan Fareed M.},
  title   = {A survey on detecting healthcare concept drift in {AI}/{ML} models from a finance perspective},
  journal = {Frontiers in Artificial Intelligence},
  volume  = {5},
  pages   = {955314},
  year    = {2023},
  doi     = {10.3389/frai.2022.955314},
}

@article{Sun2017,
  author  = {Jie Sun and Hitoshi Fujita and Peng Chen and Hongbo Li},
  title   = {Dynamic Financial Distress Prediction with Concept Drift Based on Time Weighting Combined with AdaBoost Support Vector Machine Ensemble},
  journal = {Knowledge-Based Systems},
  volume  = {120},
  pages   = {4--14},
  year    = {2017},
  doi     = {10.1016/j.knosys.2016.11.006},
}

@article{Apsemidis2020,
  author  = {Apsemidis, Alexandros and Psarakis, Stelios and Moguerza, Javier M.},
  title   = {A review of machine learning kernel methods in statistical process monitoring},
  journal = {Computers \& Industrial Engineering},
  volume  = {149},
  pages   = {106776},
  year    = {2020},
  doi     = {10.1016/j.cie.2020.106776},
}

@book{Barocas2019,
  title = {Fairness and Machine Learning: Limitations and Opportunities},
  author = {Solon Barocas and Moritz Hardt and Arvind Narayanan},
  publisher = {MIT Press},
  year = {2023}
}

@inproceedings{Wang2003,
  author    = {Haixun Wang and Wei Fan and Philip S. Yu and Jiawei Han},
  title     = {Mining Concept-Drifting Data Streams Using Ensemble Classifiers},
  booktitle = {KDD},
  pages     = {226--235},
  year      = {2003}
}

@article{Abbasi2022,
  author    = {Saddam Akber Abbasi and Ali Yeganeh and Sandile C. Shongwe},
  title     = {Monitoring non-parametric profiles using adaptive {EWMA} control chart},
  journal   = {Scientific Reports},
  volume    = {12},
  number    = {14336},
  year      = {2022},
  doi       = {10.1038/s41598-022-18381-8},
}

@article{Chang2010,
  author    = {Shing I. Chang and Srikanth Yadama},
  title     = {Statistical process control for monitoring non-linear profiles using wavelet filtering and {B}-spline approximation},
  journal   = {International Journal of Production Research},
  volume    = {48},
  number    = {4},
  pages     = {1049--1068},
  year      = {2010},
  doi       = {10.1080/00207540802454799},
}

@article{Febbo2013,
  author  = {Mariano Febbo and Sebasti{\'a}n P. Machado},
  title   = {Nonlinear dynamic vibration absorbers with a saturation},
  journal = {Journal of Sound and Vibration},
  year    = {2013},
  volume  = {332},
  number  = {6},
  pages   = {1465--1483},
  doi     = {10.1016/j.jsv.2012.11.025}
}

@article{Efron1983,
  title={Estimating the Error Rate of a Prediction Rule: Improvement on Cross-Validation},
  author={Efron, Bradley},
  journal={Journal of the American Statistical Association},
  volume={78},
  number={382},
  pages={316--331},
  year={1983},
  publisher={Taylor \& Francis}
}

@inproceedings{Yosinski2014,
  title={How transferable are features in deep neural networks?},
  author={Yosinski, Jason and Clune, Jeff and Bengio, Yoshua and Lipson, Hod},
  booktitle={Advances in Neural Information Processing Systems},
  volume={27},
  year={2014}
}

@article{Steiner2000,
  title={Monitoring surgical performance using risk-adjusted cumulative sum charts},
  author={Steiner, Stefan H and Cook, Richard J and Farewell, Vern T and Treasure, Tom},
  journal={Biostatistics},
  volume={1},
  number={4},
  pages={441--452},
  year={2000},
  publisher={Oxford University Press}
}

@article{Zhang2021,
  author    = {Zhang, Kungang and Apley, Daniel W. and Chen, Wei},
  title     = {Nonstationarity Analysis of Materials Microstructures via Fisher Score Vectors},
  journal   = {Acta Materialia},
  volume    = {211},
  pages     = {116818},
  year      = {2021},
  month     = {June},
  doi       = {10.1016/j.actamat.2021.116818},
  issn      = {1359-6454},
  publisher = {Elsevier}
}

@article{White1982,
  author  = {White, Halbert},
  title   = {Maximum Likelihood Estimation of Misspecified Models},
  journal = {Econometrica},
  volume  = {50},
  number  = {1},
  pages   = {1--25},
  year    = {1982},
  publisher = {The Econometric Society}
}

\newpage

\appendix
\section{Assumptions and Proof of Theorem 3.1}\label{app}
\renewcommand{\theequation}{B.\arabic{equation}}
\setcounter{equation}{0}

\renewcommand{\thetheorem}{B.\arabic{theorem}}
\setcounter{theorem}{0}
\renewcommand{\thelemma}{A.\arabic{lemma}}
\setcounter{lemma}{0}
\renewcommand{\theassumption}{A.\arabic{assumption}}
\setcounter{assumption}{0}
\renewcommand{\thecorollary}{A.\arabic{corollary}}


To derive the CLs presented in Section \ref{Bias Detection Algorithms and Derivations}, we state the regularity and bootstrap assumptions used throughout. These conditions are standard for MLE-based score methods and bootstrap approximations.

\subsection*{A.1 Assumptions and Preliminary Results}
\renewcommand{\theequation}{A.\arabic{equation}}
\setcounter{equation}{0}

\begin{assumption}[\textbf{Model and MLE Regularity}]\label{ass:model}
The training sample 
$\mathcal{D}=\{(\mathbf{x}_i,y_i)\}_{i=1}^n$
is drawn i.i.d.\ from a joint distribution $\mathbb{P}_0(Y,\mathbf{X})$ such that:
\begin{enumerate}[label=(\roman*)]
\item \textbf{Correct specification}:  
The supervised learning model $g_\gamma(\boldsymbol{\theta};\mathbf{x})$ induces a 
conditional density $\mathbb{P}(y\mid \mathbf{x};\boldsymbol{\theta})$ that is correctly 
specified at a true parameter $\boldsymbol{\theta}^0$.
\item \textbf{Identifiability}:  
The parameter $\boldsymbol{\theta}^0$ is an identifiable element of the parametric family.
\item \textbf{Smoothness and Fisher information}:  
The log-likelihood 
$\ell(\boldsymbol{\theta};\mathbf{x},y)=\log \mathbb{P}(y\mid \mathbf{x};\boldsymbol{\theta})$
is twice continuously differentiable in a neighborhood of $\boldsymbol{\theta}^0$,  
and the Fisher information matrix 
$\mathbf{I}(\boldsymbol{\theta}^0)
=\mathbb{E}_0[-\nabla^2_{\boldsymbol{\theta}}
 \log \mathbb{P}(Y\mid \mathbf{X};\boldsymbol{\theta}^0)]$
exists, is finite, and satisfies $\mathbf{I}(\boldsymbol{\theta}^0)\succ 0$.
\item \textbf{Score moments}:  
The score vectors 
$\mathbf{s}(\boldsymbol{\theta}^0;\mathbf{X},Y)
 = \nabla_{\boldsymbol{\theta}}
   \log \mathbb{P}(Y\mid \mathbf{X};\boldsymbol{\theta})|_{\boldsymbol{\theta}=\boldsymbol{\theta}^0}$
have finite second moments.  
\item \textbf{Asymptotic normality of the estimator}:  
The (possibly regularized) MLE $\hat{\boldsymbol{\theta}}$ satisfies
\[
\sqrt{n}\,(\hat{\boldsymbol{\theta}}-\boldsymbol{\theta}^0)
\;\xrightarrow{d}\;
\mathcal{N}\!\left(\mathbf{0},\,\mathbf{I}(\boldsymbol{\theta}^0)^{-1}\right).
\]
\end{enumerate}
\end{assumption}

\begin{assumption}[\textbf{The typical bootstrap assumption}]\label{bootstrap assumption}
If $\mathcal{D}^{\text{new}}$ is drawn from the same distribution as $\mathcal{D}$, the joint distribution of $(\mathcal{S}_{.368}^{\text {new }}, \overline{\mathbf{s}}, \widehat{\boldsymbol{\Sigma}}, \hat{\boldsymbol{\theta}})$ is the same as the joint distribution of the analogous quantities $\left(\mathcal{S}_{O O B}^b, \overline{\mathbf{s}}^b, \widehat{\boldsymbol{\Sigma}}^b, \hat{\boldsymbol{\theta}}^{b}\right)$ from the bootstrapping procedure, where $\mathcal{S}_{.368}^{\text {new }}$ is any randomly drawn subset of $\mathcal{S}^{\text {new }}$ having the same cardinality $0.368 n$ as $\mathcal{S}_{O O B}^b$.
\end{assumption}

\begin{assumption}[\textbf{Misspecification regularity}]
\label{ass:misspec}
Let $\mathbb{P}$ denote a joint distribution of $(\mathbf{X},Y)$, and let
$p_{\boldsymbol{\theta}}(y\mid\mathbf{x})$ be the conditional density represented/approximated by the supervised learning model
(possibly misspecified). Assume:
\begin{enumerate}[label=(\roman*)]
\item \textbf{Existence and uniqueness of pseudo-true parameter (See \citet{White1982} for the concept of pseudo-true parameters):}
The objective
\[
M_{\mathbb{P}}(\boldsymbol{\theta})
:= \mathbb{E}_{\mathbb{P}}
\!\left[\log p_{\boldsymbol{\theta}}(Y\mid\mathbf{X})\right]
\]
admits a unique maximizer
$\boldsymbol{\theta}_{\mathbb{P}}^\star$ (the pseudo-true parameters)
in the interior of a parameter space
$\Theta \subset \mathbb{R}^p$.

\item \textbf{Differentiation under the expectation:}
The function
$\log p_{\boldsymbol{\theta}}(y\mid\mathbf{x})$
is differentiable in $\boldsymbol{\theta}$, and
\[
\nabla_{\boldsymbol{\theta}} M_{\mathbb{P}}(\boldsymbol{\theta})
=
\mathbb{E}_{\mathbb{P}}
\!\left[
\mathbf{s}(\boldsymbol{\theta};\mathbf{X},Y)
\right],
\]
with interchangeability of gradient and expectation operators.

\item \textbf{Uniqueness of stationary point:}
For $\mathbb{P}\in\{\mathbb{P}_0,\mathbb{P}_1\}$, the equation
$\nabla_{\boldsymbol{\theta}} M_{\mathbb{P}}(\boldsymbol{\theta})=\mathbf{0}$
has the unique solution
$\boldsymbol{\theta}=\boldsymbol{\theta}_{\mathbb{P}}^\star$.
\end{enumerate}
\end{assumption}

\begin{assumption}[\textbf{Monitoring process}]\label{ass:mewma}
New observations $(\mathbf{x}_{n+i},y_{n+i})$ used for monitoring are drawn 
i.i.d.\ from the same in-control distribution $\mathbb{P}_0(Y,\mathbf{X})$ as the 
training data. Conditional on $\hat{\boldsymbol{\theta}}$, the score vectors 
$\{\mathbf{s}_{n+i}\}_{i\ge1}$ form an i.i.d.\ sequence with conditional mean 
$\boldsymbol{\mu}(\hat{\boldsymbol{\theta}})$ and conditional covariance 
$\mathbf{V}(\hat{\boldsymbol{\theta}})$, as in equation \eqref{score vector dist}.
\end{assumption}

\begin{assumption}[\textbf{Regularity of the stationary bootstrap distribution}]
\label{ass:density}
The stationary monitoring statistic $T_\infty^*$ constructed in 
Appendix~\ref{app2} has a bounded conditional density given $\mathcal D$:
\[
\sup_x\, f_\infty(x\mid\mathcal D) \;\le\; M(\mathcal D) \;<\; \infty.
\]
\end{assumption}

\begin{lemma}[\textbf{Moments of the MEWMA statistic under $\mathbb{P}_0$}]\label{lem:zn-moments}
Suppose Assumptions~\ref{ass:model} and~\ref{ass:mewma} hold, and that there is
no concept drift so that the new observations $(\mathbf{x}_{n+i},y_{n+i})$ are
drawn i.i.d.\ from $\mathbb{P}_0(Y,\mathbf{X})$.
Let $\{\mathbf{s}_{n+i}\}_{i\ge1}$ and $\{\mathbf{z}_{n+i}\}_{i\ge1}$ be the
score vectors and MEWMA statistics defined in~\eqref{score vector dist} and in
Assumption~\ref{ass:mewma}, respectively. Then, for each $i\ge1$,
\begin{equation}\label{z_moments}
\begin{aligned}
\mathbb{E}_0\!\left[\mathbf{z}_{n+i}\,\middle|\,\hat{\boldsymbol{\theta}}\right]
    &= \bigl[1-(1-\lambda)^{i}\bigr]\,
       \boldsymbol{\mu}(\hat{\boldsymbol{\theta}}),\\[4pt]
\operatorname{Cov}_0\!\left[\mathbf{z}_{n+i}\,\middle|\,\hat{\boldsymbol{\theta}}\right]
    &= \frac{\lambda}{2-\lambda}\,
       \bigl[1-(1-\lambda)^{2i}\bigr]\,
       \mathbf{V}(\hat{\boldsymbol{\theta}}),
\end{aligned}
\end{equation}
where $\boldsymbol{\mu}(\hat{\boldsymbol{\theta}})$ and
$\mathbf{V}(\hat{\boldsymbol{\theta}})$ are the conditional mean and covariance
of $\mathbf{s}_{n+i}$ given $\hat{\boldsymbol{\theta}}$ in~\eqref{score vector dist}.
\end{lemma}

\begin{proof}
By iterating the recursion
$\mathbf{z}_{n+i} = \lambda \mathbf{s}_{n+i} + (1-\lambda)\mathbf{z}_{n+i-1}$
with $\mathbf{z}_n=\mathbf{0}$, we obtain the moving-average representation
\[
\mathbf{z}_{n+i}
 = \lambda\Bigl[
    \mathbf{s}_{n+i}
   + (1-\lambda)\mathbf{s}_{n+i-1}
   + (1-\lambda)^2\mathbf{s}_{n+i-2}
   + \cdots
   + (1-\lambda)^{i-1}\mathbf{s}_{n+1}
   \Bigr].
\]
Conditional on $\hat{\boldsymbol{\theta}}$, the score vectors
$\{\mathbf{s}_{n+i}\}_{i\ge1}$ are i.i.d.\ with mean
$\boldsymbol{\mu}(\hat{\boldsymbol{\theta}})$ and covariance
$\mathbf{V}(\hat{\boldsymbol{\theta}})$ by~\eqref{score vector dist}. Taking
conditional expectations and using linearity,
\[
\mathbb{E}_0\!\left[\mathbf{z}_{n+i}\,\middle|\,\hat{\boldsymbol{\theta}}\right]
 = \lambda\sum_{k=0}^{i-1}(1-\lambda)^k\,
   \boldsymbol{\mu}(\hat{\boldsymbol{\theta}})
 = \bigl[1-(1-\lambda)^i\bigr]\,
   \boldsymbol{\mu}(\hat{\boldsymbol{\theta}}),
\]
where we used the geometric-series identity
$\sum_{k=0}^{i-1}(1-\lambda)^k = [1-(1-\lambda)^i]/\lambda$.

For the covariance, conditional independence of the $\mathbf{s}_{n+i}$ and
the same moving-average representation give
\[
\operatorname{Cov}_0\!\left[\mathbf{z}_{n+i}\,\middle|\,\hat{\boldsymbol{\theta}}\right]
 = \lambda^2\sum_{k=0}^{i-1}(1-\lambda)^{2k}\,
   \mathbf{V}(\hat{\boldsymbol{\theta}})
 = \lambda^2
   \frac{1-(1-\lambda)^{2i}}{1-(1-\lambda)^2}\,
   \mathbf{V}(\hat{\boldsymbol{\theta}}).
\]
Since $1-(1-\lambda)^2 = \lambda(2-\lambda)$, this simplifies to
\[
\operatorname{Cov}_0\!\left[\mathbf{z}_{n+i}\,\middle|\,\hat{\boldsymbol{\theta}}\right]
 = \frac{\lambda}{2-\lambda}\,
   \bigl[1-(1-\lambda)^{2i}\bigr]\,
   \mathbf{V}(\hat{\boldsymbol{\theta}}),
\]
which completes the proof.
\end{proof}

Combining Assumption~\ref{bootstrap assumption}
with the conditional distribution in~\eqref{score vector dist},
the out-of-bag (OOB) score vectors associated with outer bootstrap
replicate $b$ satisfy
\begin{equation}\label{OOB dist new}
\mathbf{s}^{\,b}_{\mathrm{OOB},i}\mid \hat{\boldsymbol{\theta}}^{\,b}
\stackrel{\text{i.i.d.}}{\sim}
\bigl(\boldsymbol{\mu}(\hat{\boldsymbol{\theta}}^{\,b}),
      \mathbf{V}(\hat{\boldsymbol{\theta}}^{\,b})\bigr),
\qquad
i=1,2,\ldots,
\end{equation}
so that $\mathcal{S}^{\,b}_{\mathrm{OOB}}$ plays the same role for
$\hat{\boldsymbol{\theta}}^{\,b}$ as an independent future sample
$\mathcal{S}^{\text{new}}$ would play for $\hat{\boldsymbol{\theta}}$.
Because a bootstrap resample contains on average $0.632n$ distinct points,
its OOB set contains approximately $0.368n$ points.
Thus, for each $b$ we define the OOB mean and covariance as
\begin{equation}\label{OOB mean new}
\bar{\mathbf{s}}^{\,b}_{\mathrm{OOB}}
  = \frac{1}{\lfloor 0.368n\rfloor}
    \sum_{i=1}^{\lfloor 0.368n\rfloor} \mathbf{s}^{\,b}_{\mathrm{OOB},i},
\end{equation}
\begin{equation}\label{OOB cov new}
\hat{\boldsymbol{\Sigma}}^{\,b}_{\mathrm{OOB}}
  = \frac{1}{\lfloor 0.368n\rfloor}
    \sum_{i=1}^{\lfloor 0.368n\rfloor}
      \bigl(\mathbf{s}^{\,b}_{\mathrm{OOB},i}
            -\bar{\mathbf{s}}^{\,b}_{\mathrm{OOB}}\bigr)
      \bigl(\mathbf{s}^{\,b}_{\mathrm{OOB},i}
            -\bar{\mathbf{s}}^{\,b}_{\mathrm{OOB}}\bigr)^{\!\top}.
\end{equation}
From \eqref{OOB dist new}, the sampling variability of the OOB mean is
\begin{equation}\label{OOB mean dist new}
\bar{\mathbf{s}}^{\,b}_{\mathrm{OOB}}
    \mid \hat{\boldsymbol{\theta}}^{\,b}
    \sim
    \Bigl(
      \boldsymbol{\mu}(\hat{\boldsymbol{\theta}}^{\,b}),
      \mathbf{V}(\hat{\boldsymbol{\theta}}^{\,b})/
      \lfloor 0.368n\rfloor
    \Bigr),
\end{equation}
and the expected OOB covariance satisfies
\begin{equation}\label{OOB cov mean new}
\mathbb{E}_0\!\left[
   \hat{\boldsymbol{\Sigma}}^{\,b}_{\mathrm{OOB}}
   \mid \hat{\boldsymbol{\theta}}^{\,b}
   \right]
  =
  \frac{\lfloor 0.368n\rfloor - 1}{\lfloor 0.368n\rfloor}\,
  \mathbf{V}(\hat{\boldsymbol{\theta}}^{\,b})
  \cong \mathbf{V}(\hat{\boldsymbol{\theta}}^{\,b}),
\end{equation}
showing that $\bar{\mathbf{s}}^{\,b}_{\mathrm{OOB}}$ and
$\hat{\boldsymbol{\Sigma}}^{\,b}_{\mathrm{OOB}}$ behave like empirical mean
and covariance based on a sample of size approximately $0.368n$ from the
same population that generates future score vectors.

With these OOB quantities in hand, the inner bootstrap loop of
Algorithm~\ref{A3} treats $\mathcal{S}^{\,b}_{\mathrm{OOB}}$ as the
population from which future score vectors are generated.
For each inner replicate $j$, a bootstrap sample
$\{\mathbf{s}^{b,j}_i\}_{i\ge1}$ is drawn with replacement from
$\mathcal{S}^{\,b}_{\mathrm{OOB}}$, and the corresponding MEWMA statistic is
defined recursively by
\begin{equation}\label{bootstrap mewma}
\mathbf{z}^{b,j}_i
  = \lambda\mathbf{s}^{b,j}_i
    + (1-\lambda)\mathbf{z}^{b,j}_{i-1},
\qquad
\mathbf{z}^{b,j}_0=\mathbf{0}.
\end{equation}
This construction mirrors exactly the MEWMA definition for genuine
future score vectors in Lemma~\ref{lem:zn-moments}, except that the
``future’’ population now has mean
$\bar{\mathbf{s}}^{\,b}_{\mathrm{OOB}}$ and covariance
$\hat{\boldsymbol{\Sigma}}^{\,b}_{\mathrm{OOB}}$, each based on only
$0.368n$ observations rather than $n$.
Consequently, the MEWMA statistic $\mathbf{z}^{b,j}_i$
inherits an additional source of variability through the randomness of
$\bar{\mathbf{s}}^{\,b}_{\mathrm{OOB}}$.
Lemma~\ref{lem:boot-zn-cond-moments} formalizes the resulting
conditional mean and covariance of $\mathbf{z}^{b,j}_i$ and shows that,
relative to the true MEWMA moments in Lemma~\ref{lem:zn-moments}, an
extra covariance term of order $[1-(1-\lambda)^i]^2/(0.368n)$ appears.

\begin{lemma}[\textbf{Conditional moments of bootstrap MEWMA statistic}]
\label{lem:boot-zn-cond-moments}
Suppose Assumption~3.2 holds, and let
$\mathbf{z}^{b,j}_i$ denote the inner-bootstrap MEWMA statistic
constructed from out-of-bag score vectors as in Algorithm~\ref{A3}.
Then, for each outer replicate $b$ and inner replicate $j$,
the conditional mean and covariance of $\mathbf{z}^{b,j}_i$
given $\hat{\boldsymbol{\theta}}^{\,b}$ satisfy
\begin{equation}\label{eq:boot-zn-conditional-moments}
\begin{aligned}
\mathbb{E}_0\!\,\bigl[\mathbf{z}_i^{b,j}\mid\hat{\boldsymbol{\theta}}^{\,b}\bigr]
  &= \bigl[1-(1-\lambda)^i\bigr]\,
     \boldsymbol{\mu}\bigl(\hat{\boldsymbol{\theta}}^{\,b}\bigr),\\[4pt]
\operatorname{Cov}_0\!\,\bigl[\mathbf{z}_i^{b,j}\mid\hat{\boldsymbol{\theta}}^{\,b}\bigr]
  &\cong \left\{\frac{\lambda}{2-\lambda}\,
          \bigl[1-(1-\lambda)^{2i}\bigr]
        + \frac{1}{0.368\,n}\,
          \bigl[1-(1-\lambda)^i\bigr]^2
        \right\}
        \mathbf{V}\bigl(\hat{\boldsymbol{\theta}}^{\,b}\bigr),
\end{aligned}
\end{equation}
for all $i \ge 1$.
\end{lemma}

\begin{proof}
Fix an outer bootstrap replicate $b$ and its fitted parameter
$\hat{\boldsymbol{\theta}}^{\,b}$.
By Assumption~\ref{bootstrap assumption} and equations \eqref{OOB dist new}–\eqref{OOB cov mean new}, the
out-of-bag score vectors $\{\mathbf{s}_{\mathrm{OOB},i}^{\,b}\}$ satisfy
\[
\mathbf{s}_{\mathrm{OOB},i}^{\,b} \mid \hat{\boldsymbol{\theta}}^{\,b}
  \stackrel{\text{i.i.d.}}{\sim}
  \bigl(\boldsymbol{\mu}(\hat{\boldsymbol{\theta}}^{\,b}),
        \mathbf{V}(\hat{\boldsymbol{\theta}}^{\,b})\bigr),
\]
and the OOB sample mean and covariance have conditional moments
\[
\mathbb{E}_0\!\bigl[\overline{\mathbf{s}}_{\mathrm{OOB}}^{\,b}
   \bigm| \hat{\boldsymbol{\theta}}^{\,b}\bigr]
  = \boldsymbol{\mu}(\hat{\boldsymbol{\theta}}^{\,b}),\qquad
\mathbb{E}_0\!\bigl[\widehat{\boldsymbol{\Sigma}}_{\mathrm{OOB}}^{\,b}
   \bigm| \hat{\boldsymbol{\theta}}^{\,b}\bigr]
  \cong \mathbf{V}(\hat{\boldsymbol{\theta}}^{\,b}),\qquad
\operatorname{Cov}_0\!\bigl[\overline{\mathbf{s}}_{\mathrm{OOB}}^{\,b}
   \bigm| \hat{\boldsymbol{\theta}}^{\,b}\bigr]
  \cong \frac{1}{0.368\,n}\,\mathbf{V}(\hat{\boldsymbol{\theta}}^{\,b}).
\]

For each inner bootstrap replicate $j$, the score vectors
$\{\mathbf{s}_i^{\,b,j}: i=1,2,\dots\}$ are drawn with replacement from
$\mathcal{S}_{\mathrm{OOB}}^{\,b}$, so conditionally on
$(\hat{\boldsymbol{\theta}}^{\,b},\mathcal{D}_{\mathrm{OOB}}^{\,b})$ they are
i.i.d.\ with mean $\overline{\mathbf{s}}_{\mathrm{OOB}}^{\,b}$ and covariance
$\widehat{\boldsymbol{\Sigma}}_{\mathrm{OOB}}^{\,b}$.
The corresponding inner-loop MEWMA statistic can be written as
\[
\mathbf{z}_{i}^{\,b,j}
  = \lambda\bigl[\mathbf{s}_{i}^{\,b,j}
      + (1-\lambda)\mathbf{s}_{i-1}^{\,b,j}
      + (1-\lambda)^2\mathbf{s}_{i-2}^{\,b,j}
      + \cdots
      + (1-\lambda)^{i-1}\mathbf{s}_{1}^{\,b,j}\bigr],
  \qquad i=1,2,\dots.
\]
Applying the law of total expectation with respect to
$\mathcal{D}_{\mathrm{OOB}}^{\,b}$ gives
\begin{equation}\label{conditionalMean}
\begin{split}
\mathbb{E}_0\!\Bigl[\mathbf{z}_{i}^{\,b,j}\,\Bigm|\,\hat{\boldsymbol{\theta}}^{\,b}\Bigr]
 &= \mathbb{E}_0\!\Bigl[
      \mathbb{E}_0\!\Bigl\{
        \lambda\bigl[\mathbf{s}_{i}^{\,b,j}
        +(1-\lambda)\mathbf{s}_{i-1}^{\,b,j}
        +\cdots
        +(1-\lambda)^{i-1}\mathbf{s}_{1}^{\,b,j}\bigr]
      \,\Bigm|\,\hat{\boldsymbol{\theta}}^{\,b},
               \mathcal{D}_{\mathrm{OOB}}^{\,b}\Bigr\}
    \,\Bigm|\,\hat{\boldsymbol{\theta}}^{\,b}\Bigr] \\
 &= \mathbb{E}_0\!\Bigl[
      \lambda\sum_{k=0}^{i-1}(1-\lambda)^{k}\,
      \overline{\mathbf{s}}_{\mathrm{OOB}}^{\,b}
      \,\Bigm|\,\hat{\boldsymbol{\theta}}^{\,b}\Bigr] = \lambda\sum_{k=0}^{i-1}(1-\lambda)^{k}\,
  \mathbb{E}_0\!\Bigl[\overline{\mathbf{s}}_{\mathrm{OOB}}^{\,b}
      \Bigm|\,\hat{\boldsymbol{\theta}}^{\,b}\Bigr] = \bigl[1-(1-\lambda)^{i}\bigr]\,
    \boldsymbol{\mu}\!\bigl(\hat{\boldsymbol{\theta}}^{\,b}\bigr),
\end{split}
\end{equation}
which is exactly the first part of equation \eqref{eq:boot-zn-conditional-moments}.
Similarly, applying the law of total covariance with respect to
$\mathcal{D}_{\mathrm{OOB}}^{\,b}$ yields
\begin{equation}\label{conditionalCov}
\begin{split}
\operatorname{Cov}_0\!\Bigl[\mathbf{z}_{i}^{\,b,j}\,\Bigm|\,\hat{\boldsymbol{\theta}}^{\,b}\Bigr]
 &= \mathbb{E}_0\!\Bigl[
      \operatorname{Cov}_0\!\Bigl(
        \lambda\bigl[\mathbf{s}_{i}^{\,b,j}
           +(1-\lambda)\mathbf{s}_{i-1}^{\,b,j}
           +\cdots
           +(1-\lambda)^{i-1}\mathbf{s}_{1}^{\,b,j}\bigr]
        \,\Bigm|\,\hat{\boldsymbol{\theta}}^{\,b},
               \mathcal{D}_{\mathrm{OOB}}^{\,b}\Bigr)
      \,\Bigm|\,\hat{\boldsymbol{\theta}}^{\,b}\Bigr] \\
 &\quad + \operatorname{Cov}_0\!\Bigl[
      \mathbb{E}_0\!\Bigl(
        \lambda\bigl[\mathbf{s}_{i}^{\,b,j}
           +(1-\lambda)\mathbf{s}_{i-1}^{\,b,j}
           +\cdots
           +(1-\lambda)^{i-1}\mathbf{s}_{1}^{\,b,j}\bigr]
        \,\Bigm|\,\hat{\boldsymbol{\theta}}^{\,b},
               \mathcal{D}_{\mathrm{OOB}}^{\,b}\Bigr)
      \,\Bigm|\,\hat{\boldsymbol{\theta}}^{\,b}\Bigr].
\end{split}
\end{equation}
Conditionally on $(\hat{\boldsymbol{\theta}}^{\,b},\mathcal{D}_{\mathrm{OOB}}^{\,b})$,
the score vectors $\mathbf{s}_{i}^{\,b,j}$ are i.i.d., so the first term becomes
\[
\mathbb{E}_0\!\Bigl[
   \lambda^2 \sum_{k=0}^{i-1}(1-\lambda)^{2k}
   \widehat{\boldsymbol{\Sigma}}_{\mathrm{OOB}}^{\,b}
   \,\Bigm|\,\hat{\boldsymbol{\theta}}^{\,b}\Bigr]
 = \lambda^2 \sum_{k=0}^{i-1}(1-\lambda)^{2k}\,
   \mathbb{E}_0\!\Bigl[\widehat{\boldsymbol{\Sigma}}_{\mathrm{OOB}}^{\,b}
   \Bigm|\,\hat{\boldsymbol{\theta}}^{\,b}\Bigr]
 \cong \frac{\lambda}{2-\lambda}\,
      \bigl[1-(1-\lambda)^{2i}\bigr]\,
      \mathbf{V}\!\bigl(\hat{\boldsymbol{\theta}}^{\,b}\bigr).
\]
The second term is the covariance of a linear combination of
$\overline{\mathbf{s}}_{\mathrm{OOB}}^{\,b}$, and using
$\operatorname{Cov}_0[\overline{\mathbf{s}}_{\mathrm{OOB}}^{\,b}
 \mid \hat{\boldsymbol{\theta}}^{\,b}]
 \cong \mathbf{V}(\hat{\boldsymbol{\theta}}^{\,b})/(0.368\,n)$ gives
\[
\operatorname{Cov}_0\!\Bigl\{
   \lambda\sum_{k=0}^{i-1}(1-\lambda)^k
   \overline{\mathbf{s}}_{\mathrm{OOB}}^{\,b}
   \,\Bigm|\,\hat{\boldsymbol{\theta}}^{\,b}\Bigr\}
 \cong \frac{1}{0.368\,n}\,
       \bigl[1-(1-\lambda)^{i}\bigr]^2\,
       \mathbf{V}\!\bigl(\hat{\boldsymbol{\theta}}^{\,b}\bigr).
\]
Combining the two contributions yields
\[
\operatorname{Cov}_0\!\Bigl[\mathbf{z}_{i}^{\,b,j}\,\Bigm|\,\hat{\boldsymbol{\theta}}^{\,b}\Bigr]
 \cong \Bigl\{
      \tfrac{\lambda}{2-\lambda}\,
      \bigl[1-(1-\lambda)^{2i}\bigr]
    + \tfrac{1}{0.368\,n}\,
      \bigl[1-(1-\lambda)^{i}\bigr]^{2}
    \Bigr\}
    \mathbf{V}\!\bigl(\hat{\boldsymbol{\theta}}^{\,b}\bigr),
\]
which is exactly the second part of equation \eqref{eq:boot-zn-conditional-moments}. This proves the stated expressions
for the conditional mean and covariance of $\mathbf{z}_i^{\,b,j}$ given
$\hat{\boldsymbol{\theta}}^{\,b}$.
\end{proof}

\subsection*{A.2 Proofs of Main Theorems}

\subsubsection*{Proof of Theorem~\ref{thm:inflation-k}}
\begin{proof}
From equations \eqref{conditionalMean} and \eqref{conditionalCov}, applying the
law of total expectation and the law of total covariance over
$\hat{\boldsymbol{\theta}}^{\,b}$ yields the unconditional mean and
covariance of $\mathbf{z}_i^{b,j}$:
\begin{equation}\label{Es}
\mathbb{E}_0\bigl[\mathbf{z}_{i}^{b,j}\bigr]
 = \mathbb{E}_0\Bigl[
     \mathbb{E}_0\bigl[\mathbf{z}_{i}^{b,j}\mid\hat{\boldsymbol{\theta}}^{\,b}\bigr]
   \Bigr]
 = \bigl[1-(1-\lambda)^{i}\bigr]\,
   \mathbb{E}_0\bigl[\boldsymbol{\mu}(\hat{\boldsymbol{\theta}}^{\,b})\bigr],
\end{equation}
and
\begin{equation}\label{Covs}
\begin{aligned}
\operatorname{Cov}_0\bigl[\mathbf{z}_{i}^{b,j}\bigr]
 &= \mathbb{E}_0\Bigl[
      \operatorname{Cov}_0\bigl[\mathbf{z}_{i}^{b,j}\mid\hat{\boldsymbol{\theta}}^{\,b}\bigr]
    \Bigr]
    + \operatorname{Cov}_0\Bigl[
      \mathbb{E}_0\bigl[\mathbf{z}_{i}^{b,j}\mid\hat{\boldsymbol{\theta}}^{\,b}\bigr]
      \Bigr]                                                              \\
 &= \left\{\frac{\lambda}{2-\lambda}\,
           \bigl[1-(1-\lambda)^{2i}\bigr]
         + \frac{1}{0.368\,n}\,
           \bigl[1-(1-\lambda)^{i}\bigr]^2
    \right\}\,
    \mathbb{E}_0\bigl[\mathbf{V}(\hat{\boldsymbol{\theta}}^{\,b})\bigr]   \\
 &\quad
    + \bigl[1-(1-\lambda)^{i}\bigr]^{2}\,
      \operatorname{Cov}_0\bigl[\boldsymbol{\mu}(\hat{\boldsymbol{\theta}}^{\,b})\bigr].
\end{aligned}
\end{equation}

To approximate the terms on the right-hand sides of equations \eqref{Es} and
\eqref{Covs} and relate them to equation \eqref{z_moments}, consider first-order
Taylor expansions of $\boldsymbol{\mu}(\hat{\boldsymbol{\theta}}^{\,b})$
and $\mathbf{V}(\hat{\boldsymbol{\theta}}^{\,b})$ about the true
parameter $\boldsymbol{\theta}^0$. Assume $\boldsymbol{\mu}(\boldsymbol{\theta})$
and $\mathbf{V}(\boldsymbol{\theta})$ are continuously differentiable in
a neighborhood of $\boldsymbol{\theta}^0$. Then
\begin{equation}\label{mu}
\boldsymbol{\mu}(\hat{\boldsymbol{\theta}}^{\,b})
 \cong \boldsymbol{\mu}(\boldsymbol{\theta}^{0})
   + \Bigl[\nabla_{\boldsymbol{\theta}}^{\!\top}
           \boldsymbol{\mu}(\boldsymbol{\theta})\Bigr]_{\boldsymbol{\theta}=\boldsymbol{\theta}^{0}}
     \bigl(\hat{\boldsymbol{\theta}}^{\,b}-\boldsymbol{\theta}^{0}\bigr).
\end{equation}
Using the standard likelihood identities at $\boldsymbol{\theta}^0$,
namely that the score has mean zero and
$
\boldsymbol{\mu}(\boldsymbol{\theta}^{0}) = \mathbf{0},$, and $
[\nabla_{\boldsymbol{\theta}}^{\!\top}
      \boldsymbol{\mu}(\boldsymbol{\theta})]_{\boldsymbol{\theta}=\boldsymbol{\theta}^{0}}
  = -\,\mathbf{I}(\boldsymbol{\theta}^{0}),
$
equation~\eqref{mu} simplifies to
\[
\boldsymbol{\mu}(\hat{\boldsymbol{\theta}}^{\,b})
 \cong -\,\mathbf{I}(\boldsymbol{\theta}^{0})
        \bigl(\hat{\boldsymbol{\theta}}^{\,b}-\boldsymbol{\theta}^{0}\bigr).
\]
Similarly,
\begin{equation}\label{V}
\mathbf{V}(\hat{\boldsymbol{\theta}}^{\,b})
 \cong \mathbf{I}(\boldsymbol{\theta}^{0})
   + \Bigl[\nabla_{\boldsymbol{\theta}}^{\!\top}
           \mathbf{V}(\boldsymbol{\theta})\Bigr]_{\boldsymbol{\theta}=\boldsymbol{\theta}^{0}}
     \bigl(\hat{\boldsymbol{\theta}}^{\,b}-\boldsymbol{\theta}^{0}\bigr),
\end{equation}
Under Assumption~\ref{ass:model}, the MLE
$\hat{\boldsymbol{\theta}}$ satisfies the standard asymptotic normality,
and the same holds for $\hat{\boldsymbol{\theta}}^{\,b}$ by
Assumption~\ref{bootstrap assumption}. Combining this with
\eqref{mu}–\eqref{V} gives the approximations
\begin{equation}\label{Emu}
\begin{aligned}
\mathbb{E}_0\left[\mathbf{\mu}\left(\hat{\boldsymbol{\theta}}^{b}\right)\right] & \cong \mathbb{E}_0\left[-\mathbf{I}\left(\boldsymbol{\theta}^{0}\right)\left(\hat{\boldsymbol{\theta}}^{b}-\boldsymbol{\theta}^{0}\right)\right] \cong \mathbf{0},
\end{aligned}
\end{equation}
\begin{equation}\label{Covmu}
\begin{aligned}
\operatorname{Cov}_0\left[\mathbf{\mu}\left(\hat{\boldsymbol{\theta}}^{b}\right)\right] \cong \operatorname{Cov}_0\left[-\mathbf{I}\left(\boldsymbol{\theta}^{0}\right)\left(\hat{\boldsymbol{\theta}}^{b}-\boldsymbol{\theta}^{0}\right)\right] \cong \mathbf{I}\left(\boldsymbol{\theta}^{0}\right) \frac{\mathbf{I}^{-1}(\boldsymbol{\theta}^{0})}{n} \mathbf{I}\left(\boldsymbol{\theta}^{0}\right) = \frac{\mathbf{I}(\boldsymbol{\theta}^{0})}{n},
\end{aligned}
\end{equation}
and
\begin{equation}\label{EV}
\mathbb{E}_0\left[\mathbf{V}\left(\hat{\boldsymbol{\theta}}^{b}\right)\right] \cong \mathbb{E}_0\left[\mathbf{I}\left(\boldsymbol{\theta}^{0}\right)+\left.\nabla_{\boldsymbol{\theta}}^{\!\top}\mathbf{V}(\boldsymbol{\theta}^{0})\right|_{\boldsymbol{\theta}=\boldsymbol{\theta}^{0}}\left(\hat{\boldsymbol{\theta}}^{b}-\boldsymbol{\theta}^{0}\right)\right] = \mathbf{I}\left(\boldsymbol{\theta}^{0}\right).
\end{equation}
where the term involving
$\nabla_{\boldsymbol{\theta}}^{\!\top}\mathbf{V}(\boldsymbol{\theta}^{0})$
drops out in the expectation because
$\mathbb{E}_0[\hat{\boldsymbol{\theta}}^{\,b}-\boldsymbol{\theta}^{0}]\approx\mathbf{0}$.

Substituting \eqref{Emu}–\eqref{EV} into \eqref{Es}–\eqref{Covs} yields
\begin{equation}\label{sampling_es and cov}
\mathbb{E}_0\bigl[\mathbf{z}_{i}^{b,j}\bigr]
 \cong \mathbf{0},
\qquad
\operatorname{Cov}_0\bigl[\mathbf{z}_{i}^{b,j}\bigr]
 \cong \left\{\frac{\lambda}{2-\lambda}\,
                \bigl[1-(1-\lambda)^{2i}\bigr]
              + \frac{3.72}{n}\,
                \bigl[1-(1-\lambda)^{i}\bigr]^2
        \right\}\mathbf{I}(\boldsymbol{\theta}^{0}).
\end{equation}
Using \eqref{z_moments} and repeating the same Taylor and asymptotic
arguments for $\hat{\boldsymbol{\theta}}$ (rather than
$\hat{\boldsymbol{\theta}}^{\,b}$) gives the analogous result for
$\mathbf{z}_i$ (now without the OOB mean-variability term),
\begin{equation}
\mathbb{E}_0\bigl[\mathbf{z}_{i}\bigr]
 \cong \mathbf{0},
\qquad
\operatorname{Cov}_0\bigl[\mathbf{z}_{i}\bigr]
 \cong \left\{\frac{\lambda}{2-\lambda}\,
                \bigl[1-(1-\lambda)^{2i}\bigr]
              + \frac{1}{n}\,
                \bigl[1-(1-\lambda)^{i}\bigr]^2
        \right\}\mathbf{I}(\boldsymbol{\theta}^{0}).
\end{equation}

Thus the unconditional covariance matrices of
$\mathbf{z}^{b,j}_i$ and $\mathbf{z}_i$ differ only by a scalar factor
\[
\operatorname{Cov}_0\bigl[\mathbf{z}^{b,j}_i\bigr]
 \cong k(\lambda,i,n)\,
        \operatorname{Cov}_0\bigl[\mathbf{z}_i\bigr],
\]
where $k(\lambda,i,n)$ is given by \eqref{eq:k-def-short}. This is
exactly the relationship stated in Theorem~\ref{thm:inflation-k}.
\end{proof}

\section{Stabilization of the CL}\label{app2}
\renewcommand{\theequation}{B.\arabic{equation}}
\setcounter{equation}{0}

Throughout this section, $\mathcal{D}$ denotes the fixed Phase~I training
sample. For each monitoring time $i\ge 1$, Algorithm~\ref{A3} defines a
two-stage resampling scheme that, given $\mathcal{D}$, induces a distribution
over bootstrap monitoring statistics. Let $T_i^*$ denote a generic bootstrap
replicate of the monitoring statistic at time $i$, and let
$F_i(\cdot \mid \mathcal{D})$ denote its conditional distribution function,
accounting for both Phase~I estimation uncertainty (due to bootstrap sampling
of the training data from $\mathcal{D}$) and Phase~II variability.

In this appendix we show that the CL for our control chart stabilizes
(converges to a constant as $i \to \infty$) by showing that
$F_i(\cdot\mid\mathcal{D})$ converges to a limiting distribution
$F_\infty(\cdot\mid\mathcal{D})$ as $i \to \infty$.

\begin{theorem}[\textbf{Distributional stabilization of the bootstrap
monitoring statistic}]
\label{thm:cdf-stabilization}
Consider a fixed Phase~I sample $\mathcal{D}$ such that each
$\widehat{\boldsymbol{\Sigma}}^b$ is positive definite almost surely
over the outer bootstrap draws. Then the bootstrap conditional distributions
$F_i(\cdot\mid\mathcal{D})$ converge weakly to a limit
$F_\infty(\cdot\mid\mathcal{D})$ as $i\to\infty$. If, in addition,
Assumption~\ref{ass:density} holds, then the convergence is uniform:
\begin{equation}\label{eq:cdf-stab}
\sup_x\,\bigl|F_i(x\mid\mathcal{D})-F_\infty(x\mid\mathcal{D})\bigr|
\;\to\; 0.
\end{equation}
\end{theorem}

\begin{proof}
Fix the Phase~I training sample $\mathcal{D}$ and condition on $\mathcal{D}$
throughout. Recall that, given $\mathcal{D}$, the only sources of randomness
in $T_i^{b,j}$ are the outer bootstrap mechanism (which determines
$\mathcal{D}^b$, $\hat{\boldsymbol{\theta}}^b$, $\overline{\mathbf{s}}^b$,
$\widehat{\boldsymbol{\Sigma}}^b$, and $\mathcal{S}_{\mathrm{OOB}}^b$) and
the inner bootstrap mechanism (which draws
$\{\mathbf{s}_i^{b,j}\}_{i\ge 1}$ i.i.d.\ from $\mathcal{S}_{\mathrm{OOB}}^b$).

Fix an outer bootstrap replicate $b$ and temporarily condition additionally on
$\mathcal{D}^b$. Under this double conditioning, $\overline{\mathbf{s}}^b$,
$\widehat{\boldsymbol{\Sigma}}^b$, and $\mathcal{S}_{\mathrm{OOB}}^b$ are
all fixed, and the inner bootstrap score vectors
$\{\mathbf{s}_i^{b,j}\}_{i\ge 1}$ are i.i.d.\ draws from the fixed finite
empirical distribution of $\mathcal{S}_{\mathrm{OOB}}^b$, which has finite
second moment. Because the MEWMA recursion is a stable linear filter
($|1-\lambda|<1$) applied to these i.i.d.\ draws, standard theory for
stationary EWMA processes implies that the distribution of
$\mathbf{z}_i^{b,j}$ converges weakly to a stationary limiting distribution
as $i\to\infty$, conditioned on $(\mathcal{D},\mathcal{D}^b)$. Since
$k(\lambda,i,n)\to k_\infty(\lambda,n):=\lim_{i\to\infty}k(\lambda,i,n)$
is a convergent scalar sequence and $\widehat{\boldsymbol{\Sigma}}^b$ is
positive definite, the continuous mapping theorem implies that the
distribution of $T_i^{b,j}$ (conditioned on $(\mathcal{D},\mathcal{D}^b)$)
likewise converges weakly to a limiting distribution. 

Denote the conditional CDFs of $T_i^{b,j}$ and $T_\infty^{b,j}$ by
$F_i^b(\cdot\mid\mathcal{D},\mathcal{D}^b)$ and $F_\infty^b(\cdot\mid\mathcal{D},\mathcal{D}^b)$, respectively.
Marginalizing over $\mathcal{D}^b$ and applying the bounded convergence
theorem (since CDFs are bounded between 0 and 1),
\[
F_i(x\mid\mathcal{D})
  = \mathbb{E}\!\left[
      F_i^b(x\mid\mathcal{D},\mathcal{D}^b)\mid\mathcal{D}
    \right]
  \;\to\;
  \mathbb{E}\!\left[
      F_\infty^b(x\mid\mathcal{D},\mathcal{D}^b)\mid\mathcal{D}
    \right]
  =: F_\infty(x\mid\mathcal{D})
\]
at every continuity point of $F_\infty(\cdot\mid\mathcal{D})$, establishing
weak convergence
$F_i(\cdot\mid\mathcal{D})\Rightarrow F_\infty(\cdot\mid\mathcal{D})$.
Under Assumption~\ref{ass:density}, $F_\infty(\cdot\mid\mathcal{D})$ has a
bounded density and is therefore continuous, so P\'{o}lya's theorem upgrades
weak convergence to uniform convergence of the CDFs, yielding~\eqref{eq:cdf-stab}.
\end{proof}

The uniform CDF bound~\eqref{eq:cdf-stab} directly implies that every
feature of the bootstrap distribution, including its quantiles, converges,
provided the relevant functional of the CDF is continuous. In particular,
the control limit $\mathrm{CL}_i$ converges because it is a quantile.

\renewcommand{\thetheorem}{C.\arabic{theorem}}
\setcounter{theorem}{0}
\section{Detectability of A Change under Model Misspecification}\label{app25}
\renewcommand{\theequation}{C.\arabic{equation}}
\setcounter{equation}{0}


In this section we investigate whether the score-based monitoring mechanism
can detect changes in the predictive relationship when the supervised
learning model is misspecified. In particular, we focus on whether a
distributional shift induces a nonzero mean in the monitored score vector,
which is the fundamental signal driving detection.

Let $\mathbb{P}_0$ denote the in-control joint distribution of
$(\mathbf{X},Y)$, and let $p_{\boldsymbol{\theta}}(y\mid\mathbf{x})$ denote
the conditional distribution represented by the supervised learning model
(possibly misspecified), where $p_{\boldsymbol{\theta}}(y\mid\mathbf{x})$
is interpreted as a probability density function when $Y$ is continuous and
as a probability mass function when $Y$ is categorical.
Define the pseudo-true parameter
\[
\boldsymbol{\theta}^\star
:= \arg\max_{\boldsymbol{\theta}\in\Theta}\,
\mathbb{E}_{\mathbb{P}_0}\!\left[\log
p_{\boldsymbol{\theta}}(Y\mid\mathbf{X})\right],
\]
and let
$\mathbf{s}(\boldsymbol{\theta};\mathbf{x},y)
:= \nabla_{\boldsymbol{\theta}}\log p_{\boldsymbol{\theta}}(y\mid\mathbf{x})$
denote the corresponding score vector.
For a post-change distribution $\mathbb{P}_1 \neq \mathbb{P}_0$, define
$\boldsymbol{\theta}_1^\star$ analogously.

\begin{theorem}[Detectability under model misspecification]
\label{thm:misspec}

Assume that the pseudo-true parameters
$\boldsymbol{\theta}^\star$ and $\boldsymbol{\theta}_1^\star$
are uniquely defined and satisfy the regularity conditions of
Assumption~\ref{ass:misspec}. Then
\begin{enumerate}[label=(\alph*)]
\item In-control score mean
$\mathbb{E}_{\mathbb{P}_0}
\!\left[
\mathbf{s}(\boldsymbol{\theta}^\star;\mathbf{X},Y)
\right]
= \mathbf{0}.$
\item
If $\boldsymbol{\theta}_1^\star \neq \boldsymbol{\theta}^\star$, then
$\mathbb{E}_{\mathbb{P}_1}
\!\left[
\mathbf{s}(\boldsymbol{\theta}^\star;\mathbf{X},Y)
\right]
\neq \mathbf{0}.$
\item
Under $\mathbb{P}_1$, the MEWMA statistic constructed from
$\mathbf{s}(\boldsymbol{\theta}^\star;\mathbf{X},Y)$
converges geometrically in mean to a nonzero vector.
\end{enumerate}
\end{theorem}

\begin{proof}[Proof of Theorem~\ref{thm:misspec}]
Parts~(a) and~(b) are direct consequences of the theory of pseudo-true
parameters under model misspecification developed by \citet{White1982}.
Specifically, part~(a) follows from the first-order optimality condition
$\nabla_{\boldsymbol{\theta}} M_{\mathbb{P}_0}(\boldsymbol{\theta}^\star)
= \mathbf{0}$ combined with the interchange of gradient and expectation
(Assumption~\ref{ass:misspec}(ii)), which gives
$\mathbb{E}_{\mathbb{P}_0}[\mathbf{s}(\boldsymbol{\theta}^\star;
\mathbf{X}, Y)]
= \nabla_{\boldsymbol{\theta}} M_{\mathbb{P}_0}(\boldsymbol{\theta}^\star)
= \mathbf{0}$.
Part~(b) follows because
$\mathbb{E}_{\mathbb{P}_1}[\mathbf{s}(\boldsymbol{\theta}^\star;
\mathbf{X}, Y)]
= \nabla_{\boldsymbol{\theta}} M_{\mathbb{P}_1}(\boldsymbol{\theta}^\star)$,
and if this were zero then $\boldsymbol{\theta}^\star$ would be a stationary
point of $M_{\mathbb{P}_1}$, hence equal to
$\boldsymbol{\theta}_1^\star$ by the uniqueness assumption
(Assumption~\ref{ass:misspec}(iii)), contradicting
$\boldsymbol{\theta}_1^\star \neq \boldsymbol{\theta}^\star$.
Part~(c) is elementary: under $\mathbb{P}_1$, the score vectors
$\{\mathbf{s}_i\}_{i \ge 1}$ are i.i.d.\ with mean
$\boldsymbol{\mu}_1
:= \mathbb{E}_{\mathbb{P}_1}[\mathbf{s}(\boldsymbol{\theta}^\star;
\mathbf{X}, Y)] \neq \mathbf{0}$.
Iterating the MEWMA recursion gives
$\mathbb{E}_{\mathbb{P}_1}[\mathbf{z}_i]
= [1-(1-\lambda)^i]\,\boldsymbol{\mu}_1
\to \boldsymbol{\mu}_1 \neq \mathbf{0}$
at geometric rate $(1-\lambda)$.
\end{proof}
Theorem~\ref{thm:misspec} shows that detectability of concept drift does not
rely on correct model specification.
That is, any shift that changes the pseudo-true parameter produces a nonzero
mean in the monitored score vector, yielding a detectable signal in the MEWMA
statistic.
\section{Computational Details and Additional Validation}
\label{app:compute}

This appendix provides additional computational details and validation results
for the nonlinear oscillator example in Section~\ref{Nonlinear Oscillator System},
with a particular focus on pointwise false-alarm rate (PFAR) control and
post-change detection behavior under varying noise levels.

\subsection{Monte Carlo validation of PFAR}\label{app3}

We conducted an independent Monte Carlo study to assess PFAR control for the
nonlinear oscillator example.
For each configuration, we generated $R=8000$ independent monitoring streams
from the in-control data-generating mechanism and recorded whether the monitoring
statistic exceeded its corresponding time-varying control limit at each monitoring
index $i$.
The empirical PFAR at time $i$ was computed as the fraction of runs that signaled
at that time.

We considered two noise levels, $\sigma = 0.03$ (low noise) and
$\sigma = 0.20$ (high noise), using identical monitoring parameters
$\lambda = 0.01$ and nominal false-alarm rate $\alpha = 0.001$.
Control limits were computed using the proposed nested bootstrap procedure with
$B_O = 50$ outer bootstrap replicates and $B_I = 200$ inner bootstrap replicates.
For each outer bootstrap replicate, the neural network was refit for 500 training epochs.

Figure~\ref{fig:oscillator_pfar} shows the resulting PFAR
curves as functions of the monitoring index $i$ for the low- and high-noise settings,
respectively.
During the in-control period ($i \le 200$), the empirical PFAR remains close to the
nominal level $\alpha = 0.001$ in both noise regimes, demonstrating accurate
Type~I error control even under severe model misspecification.
After the change-point, the same curves reflect the pointwise probability of detection
under concept drift, which rapidly increases towards one, indicating strong eventual
detectability.

\subsection{Computational environment and computational expense}

All experiments were run on a shared high-performance computing cluster, with each
task allocated 8 CPU cores and 32\,GB of memory (Intel Xeon Gold 6338, 2.00\,GHz).
No GPU resources were used.
All computations were performed in Python~3.10.18 using PyTorch configured to run on CPU.

We distinguish between two types of computational cost.
The first is the control-limit (CL) setup time, which is incurred once per model update
and represents the primary runtime cost faced by practitioners.
The second is the monitoring cost per observation, which is negligible and therefore
not tabulated.

For the nonlinear oscillator example, the serial implementation of the proposed method
required approximately \textbf{3.2} minutes to compute the bootstrap control limits.
We additionally parallelized the outer bootstrap loop across 10 array tasks, each handling
approximately 5 outer bootstrap replicates and using 8 CPU cores.
Under this parallelization scheme, the estimated CL setup time based on the slowest
task chunk was approximately \textbf{0.7} minutes, demonstrating substantial speedup
due to the trivially parallelizable structure of the outer bootstrap.
A computational lower bound under full parallelization (one worker per outer replicate)
is approximately \textbf{0.1} minutes.

For comparison, the two-sample control-limit method of \citet{Zhang2023} has a comparable
CL setup cost, but---as demonstrated in Section~\ref{Numerical Illustrations}---exhibits poor
false-alarm calibration in small-sample and small-$\alpha$ settings.

\begin{table}[htbp]
\centering
\caption{Control-limit (CL) setup time comparison (minutes) for the nonlinear oscillator
example.
``Ours--Serial'' reports single-process CL setup time.
``Ours--Parallel (conserv)'' is a conservative estimate based on the slowest task chunk
under partial parallelization across 10 array tasks, and
``Ours--Parallel (best)'' is a computational lower bound assuming full parallelization of
outer bootstrap replicates.
Timing for \citet{Zhang2023} reflects CL setup only. Despite the nested bootstrap 
structure, the parallelized setup time of the proposed method is comparable to or faster 
than that of \citet{Zhang2023}, while providing substantially more accurate false-alarm 
control (Section~\ref{Numerical Illustrations}).}
\label{tab:compute_compare}
\small
\setlength{\tabcolsep}{4pt}
\begin{tabular}{lrrrr}
\toprule
Noise $\sigma$ & Ours-Serial (min) & Ours-Parallel (conserv) (min) & Ours-Parallel (best) (min) & Zhang et al.\ (min) \\
\midrule
0.03 & 3.2 & 0.7 & 0.1 & 0.5 \\
0.20 & 3.2 & 0.7 & 0.1 & 0.7 \\
\bottomrule
\end{tabular}
\end{table}

\end{document}